\documentclass[journal]{IEEEtran}
\usepackage{amsmath}
\usepackage{color}
\usepackage{array}
\usepackage{stfloats}
\usepackage{amsthm} 
\usepackage{amssymb}
\usepackage{float}

\usepackage{nccmath}
\usepackage{graphicx}
\usepackage[linesnumbered,ruled]{algorithm2e}
\usepackage{setspace}
\usepackage{booktabs}
\usepackage{subcaption}
\usepackage{mwe}
\usepackage{cite}
\usepackage{amsmath,amssymb,amsfonts}
\usepackage{graphicx}
\usepackage{textcomp}
\usepackage{xcolor}
\newtheorem{theorem}{Theorem}
\newtheorem{definition}{Definition}
\usepackage{algorithmicx}
\usepackage{xspace}
\newcommand{\BfPara}[1]{{\noindent\bf#1.}\xspace}

\renewcommand\baselinestretch{1.00}

\begin{document}
\title{Auction-based Charging Scheduling with Deep Learning Framework for Multi-Drone Networks}

\author{MyungJae Shin, 
	Joongheon Kim,~\IEEEmembership{Senior Member,~IEEE,}
    and Marco Levorato,~\IEEEmembership{Member,~IEEE}
    \thanks{This research was supported by the Chung-Ang University Graduate Research Scholarship in 2018 (for MyungJae Shin) and also by Institute for Information \& Communications Technology Promotion (IITP) grant funded by the Korea government (MSIT) (No.2018-0-00170, Virtual Presence in Moving Objects through 5G).}
    \thanks{Copyright (c) 2015 IEEE. Personal use of this material is permitted. However, permission to use this material for any other purposes must be obtained from the IEEE by sending a request to pubs-permissions@ieee.org.}
	\thanks{M. Shin and J. Kim are with the School of Computer Science and Engineering, Chung-Ang University, Seoul, Korea e-mails: mjshin.cau@gmail.com, joongheon@cau.ac.kr.}
	\thanks{M. Levorato is with the Department of Computer Science, Donald Bren School of Information and Computer Sciences, University of California at Irvine, Irvine, CA 92697, USA e-mail: levorato@uci.edu}
	\thanks{J. Kim is the corresponding author.}
}

\maketitle

\begin{abstract}
State-of-the-art drone technologies have severe flight time limitations due to weight constraints, which inevitably lead to a relatively small amount of available energy. Therefore,
frequent battery replacement or recharging is necessary in applications such as delivery, exploration, or support to the wireless infrastructure. Mobile charging stations (i.e., mobile stations with charging equipment) for outdoor ad-hoc battery charging is one of the feasible solutions to address this issue. However, the ability of these platforms to charge the drones is limited in terms of the number and charging time. This paper designs an auction-based mechanism to control the charging schedule in multi-drone setting. In this paper, charging time slots are auctioned, and their assignment is determined by a bidding process. The main challenge in developing this framework is the lack of prior knowledge on the distribution of the number of drones participating in the auction. Based on optimal second-price-auction, the proposed formulation, then, relies on deep learning algorithms to learn such distribution online. Numerical results from extensive simulations show that the proposed deep learning-based approach provides effective battery charging control in multi-drone scenarios.
\end{abstract}

\IEEEpeerreviewmaketitle

\section{Introduction}
The possibility to use commercial drones in a broad range of applications is being extensively studied by the research community, and they are expected to manned operations in remote locations~\cite{park2017battery}. 
In general, commercial drones have inherent limitations in the amount of energy available to support their operations. This is due to the energy/weight ratio of current energy storage technologies, where increasing the capacity of the battery beyond a certain point degrades flight time due to excessive weight.

As a consequence, effective battery management is one of the main enablers of practical deployments of drone-based technologies and applications.
Importantly, in applications requiring extensive flight time, the energy constraint problem can not be solved by only optimizing power consumption. Thus, charging during the completion of long-term tasks has been proposed to extend the operational range of the drones~\cite{park2017battery, couture2009adaptive}.
There are several ways to powering the drones which have been proposed in the literature. We can divide them into two main classes: (i) harvesting energy directly from the surrounding environment, and (ii) taking energy from an electrical source such as a charging station~\cite{couture2009adaptive}. Within the latter class of approaches, the charging stations can be either stationary or mobile. However, solutions based on stationary charging stations may constrain the geographical area of operations around specific locations. In order to deal with this issue, mobile charging stations can be used although they face other challenges~\cite{couture2009adaptive}.
As they are mobile, the size of these charging stations needs to be comparably smaller to that of fixed stations. As a consequence, the capacity of the system has limitations, leading to relatively low chargin speeds and a relatively smaller number of drones that can be charged simultaneously~\cite{couture2009adaptive,frankenberger2017mobile}. 

\begin{figure}
    \centering
    \includegraphics[width=0.7\columnwidth]{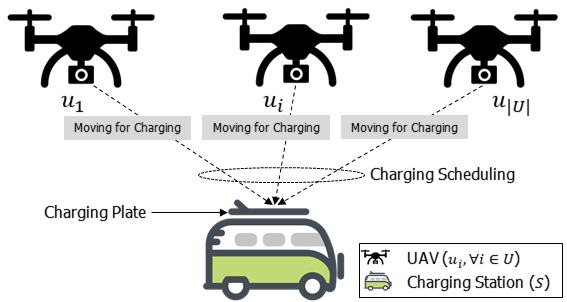}
    \caption{Multi-drone network model for mobile charging stations.}
    \label{fig:netsysmodel}
    \vspace{-6mm}
\end{figure}

Motivated by this compelling problem, we consider a scenario where multiple drones compete to access the services provided by a mobile charging station (see Fig.~\ref{fig:netsysmodel}). The framework proposed in this paper controls the charging process of the drones, where the charging station takes the role of leader in the distributed drone-charging system, and coordination within the system is supported by
Internet-of-Vehicle (IoV) networking functions~\cite{wang2018internet, hou2016vehicular,chen2018capacity}.

We take an econometric approach, where the problem of controlling the scheduling is formulated as an auction, whose objective is to maximize the utility of the drones (i.e., the difference between payment and bid during auction computation) as well as the station's revenue (i.e., payment received by the drones through charging scheduling).
In general auction problems, buyers (the drones in this system) bid to access services periodically auctioned by a seller (the mobile charging station in the considered setting).
The value of the bid is individually, and privately, estimated by each drone based on the urgency of its charging needs.
The auction approach is especially useful when there is no accurate estimation of the buyer's true valuation, and buyers are not aware of the private true values of other buyers. 
In the drone network model considered herein, the drones are assumed to be non-cooperative, that is, they operate independently and distributely. Furthermore, the mobile charging station is not assumed to know the exact true values associated with each drone, which became available only when the actual values are submitted. The auction approach we take is especially suitable to solve the problem of assigning time slots to drones in this information-limited system. 
Among the various auction formulations available (e.g., ascending auction, descending auction, first price auction, second price auction), we choose a second price auction formulation, where the highest bidder wins but the price paid is set to the second highest bid. One of the main benefits of the second price auction is that it results in a truthful auction process.

In the considered system model, the mobile station is the auctioneer and owner/seller of the resource (that is, the charging time slot) and each drone is considered as a buyer.
The drones are in competition for scheduling battery charging with price bidding via its own private valuation for auction.
As auctioneer, the mobile station (i) receives all bids from the drones, (ii) calculates the charging time allocation probabilities and payments, (iii) assigns the charging time to the drone (i.e., the winner in auction) who bids the highest value, corresponding to the largest allocation probability, (iv) announces the value which should be paid by the winner drone, and (v) receives the payment.

During the auction, drones strategically submit bids to increase their profits, i.e., utility. Similarly, the resource-owned auctioneer is not a sacrificial seller, thus it is required to consider the revenue in auctioneer, i.e., profitable. Therefore, revenue-optimal auctions have been considered as one of major objectives in auction design. 
Although there are many variants already available in the literature auction theory, the problem of simultaneously optimizing auctioneer's revenue and buyers utility is still open~\cite{yong2015double,chang2010auction,wen2015quality, yi2015multi}. Among various auction algorithms, \textit{Myerson auction} is one of the most efficient revenue-optimal single-item auctions~\cite{myerson1981}. The auction transforms the bid value, and then winner and payment is determined based on the transformed bid. At that point, if the transformation function is monotonic, the revenue-optimal auction is configured. 
Therefore, the proposed auction designs the revenue-optimal auction based on the concept of the Myerson auction. 

However, it is difficult to apply the existing auction as it is in the distributed drone network environment considered in this paper. The charging scheduling system of the drones is still in the early stages of research; and key properties of the system such as drones location distribution and residual energy distribution have not been fully characterized in the literature. 
Therefore, a system that can extract the desired data (i.e., distribution of drones), from the actual system without prior knowledge or assumptions is desirable. Therefore, this paper takes advantage of deep learning to learn important features on-the-fly from the operating environment.
Recently, frameworks combining game theory and deep learning have been active subject of research~\cite{subba2018game, tian2018application}. Results illustrate applications of such approach in various domains~\cite{cst2018wang, you2017deep, sill1998monotonic}. 
The key is that deep learning can automatically extract and learn important features from data, and it has been widely demonstrated that neural network structures can approximate complex non-linear functions~\cite{csaji2001,dutting2017, luong2017}. In this paper, we use this feature to approximate some key -- monotonic -- functions governing the behavior of the system using relatively simple neural networks~\cite{sill1998monotonic}.
Specifically, we use deep learning to learn the features necessary for the virtual transformation step of Myerson auctions. Then, the proposed auction is configured by replacing the trained deep learning network with a virtual transformation function. We remark that the functions to be learn by the deep learning layer is non-decreasing monotonic~\cite{myerson1981}.

The proposed deep learning network uses the \texttt{ReLU} (activation function) and \texttt{softmax} (classification function) which are widely used in optimization procedures. In addition, due to the fact that the operations mostly amount to linear multiplications, the proposed approach has low complexity, and its execution takes a limited amount of time. 

\BfPara{Contributions} 
Our proposed auction-based charging scheduling algorithm makes the following contributions. 
    First, the revenue of auctioneer is considered even if the drones submit false/fake bids, i.e., thus the proposed algorithm is self-configurable and truthful. 
    The proposed auction automatically learns environmental features. In distributed drone scenarios, various time varying features exist that make self-configurable nature essential to adapt to different scenarios and environments. 
    The proposed deep learning based auction structure is simple to be implemented and imposes a small computation burden. 

\BfPara{Organization} 
The rest of this paper is organized as follows.
Sec.~\ref{sec:sec2} discusses related work, and then Sec.~\ref{sec:sec3} describes the auction-based mobile charging model. In Sec.~\ref{sec:sec4_revision}, the deep learning based approach is presented. In Sec.~\ref{sec:sec4}, performance evaluation results are presented. Sec.~\ref{sec:sec5} concludes the paper. 

\section{Related Work}\label{sec:sec2}

There have been several research results to solve limited-battery and limited resource scheduling problems through auctions~\cite{park2017battery,couture2009adaptive,wellman2001auction,parkes2001auction}.
The method in \cite{park2017battery} aims at optimizing battery assignment and drone scheduling, assuming that the battery can be quickly replaced.
The joint assignment and scheduling problem is formulated as a two-stage problem, where the assignment problem is solved by a heuristic and the scheduling problem is formulated as an integer-linear programming (ILP) problem. This paper proposes the scheduling algorithm based on auction. The proposed method uses information provided by drones capable of communicating with mobile charging stations to overcome the inability of a central service provider to acquire perfect state information in a distributed drone network. However, a solution based on battery assignment necessarily maps to a stationary service station. This imposes some limitations ~\cite{couture2009adaptive}, which are mitigated when using mobile charging stations. 

In \cite{couture2009adaptive}, a systems of mobile robots executing a transportation task supported by a charging station is considered. 
The location of the charging station is a major factor in determining the operations and performance of the robots, and the paper assumes that the mobile charging station is itself an autonomous robot that attempts to incrementally improve its location. Although this work considers a mobile charging station, the problem of charging scheduling is not considered.
 In a more general scenario, the resources of the charging station are limited and the number of robots to be charged may be larger than the actual charging capacity of the station. Therefore, the charging system will need to implement forms of prioritization to optimize the charging process. The method proposed herein incorporates a notion of priority using an auction formulation based on the valuation of the drones.

In \cite{wellman2001auction}, an auction mechanism is proposed to solve a resource allocation problem in a distributed computing system. The inherently distributed nature of the system makes the resolution of the problem much harder. The paper proposes an auction-based solution to address such challenge. The proposed mechanism is configured as a two auction mechanism, used to compute optimal solutions at the single unit within the distributed scheduling problem in a computationally efficient manner. However, \cite{wellman2001auction} assumes prior knowledge of the environment where the auction mechanism is executed, which may limit its application in real-world distributed scenarios. The method we propose herein uses an auction-based solution to solve the resource allocation problem, and employs deep learning to extract the required features automatically from the environment, so that prior knowledge is not necessary.

The method in \cite{parkes2001auction} addresses a distributed train scheduling problem using an auction method. The determination of the winner is formulated as a mixed-integer problem. The bidding strategy of the buyers is solved via dynamic programming. In the proposed method, the auctioneer computes the set of bids that maximizes revenue. Both the method proposed in \cite{parkes2001auction} and the one proposed in this paper are based on an auction formulation to effectively solve resource scheduling in distributed environments and maximize the revenue of the auctioneer. However, the method in \cite{parkes2001auction} differs from the proposed deep learning based auction in terms of the required prior information to conduct the auction. The deep learning-based auction proposed in this paper only requires limited information since as it can learn in real-time environmental characteristics and parameters.

\section{Charging Scheduling Mechanism Design}\label{sec:sec3}
\setlength{\textfloatsep}{0pt} 
\begin{table}[t!]
\footnotesize
\caption{Notations}
\label{tab:param}
\begin{center}
    \scalebox{1}{
	\begin{tabular}{c|l}
    \toprule
    Variables & Descriptions \\
    \midrule [1.0pt]
$U$ & The number of drones \\ 
$\mathcal{S}$ & Mobile charging station \\
$\boldsymbol{\mathcal{B}}$ & Bid profiles\\
$\boldsymbol{\mathcal{B}_t}$ & $t$-th bid profile\\
$u_i$ & $i$-th user\\
$c_i$ & Maximum battery capacity of $u_i$\\
$r_i$ & Remaining battery capacity of $u_i$\\
$e_i$ & Average amperage draw of $u_i$\\
$h_i$ & Battery discharge of $u_i$\\
$f$ & Charging rate per unit time\\
$t_i$ & Scheduled charging time to $u_i$\\
$q_i$ & Amount of energy charged of $u_i$\\
$l_i$ & Flight time with current battery of $u_i$ \\
$v_i$ & The valuation of $u_i$\\
$b_i$ & The bid of $u_i$\\
$\overline{b_i}$ & The transformed bid of $u_i$\\
$g_i$ & Allocation probability of $u_i$\\
$\overline{p_i}$ & The virtual payment of $u_i$\\
$p_i$ & Actual payment of $u_i$\\
$\phi_i$ & The forward transformation function for $u_i$\\
$w^i_{g,n}$ & Weight of $g$-th group, $n$-th unit for $u_i$\\
$w^{shared}_{g,n}$ & Weight of $g$-th group, $n$-th unit of $phi^{shared}$\\
$\beta^i_{g,n}$ & Bias of $g$-th group, $n$-th unit for $u_i$\\
$\beta^{shared}_{g,n}$ & Bias of $g$-th group, $n$-th unit of $phi^{shared}$\\
$u_i$ & The utility of $u_i$\\
$\mathcal{G}$ & The number of groups in network\\
$\mathcal{N}$ & The number of units in group\\
$\mathcal{R}$ & The number of epoch \\
$\mathcal{T}$ & The number of bid sets \\[0.2ex]
    \bottomrule
	\end{tabular}
	}
\label{tab:param}
\end{center}
\end{table}

\BfPara{Drone Network Model} 
The system is composed of the mobile charging station $S$ and $U$ drones\footnote{The notation used in this paper is summarized in Table~\ref{tab:param}.}. The mobile station is governed by the charging service controller; and the service controller collects revenue by providing charging services. The revenue of the charging service controller is recorded and will be requested later to be paid to drone operators.
This paper assumes that the mobile station can provide charging service to only one drone in each time slot. Thus, drones competes to obtain charging opportunities.
Note that we consider a short-range Internet of Vehicles (IoV) multi-drone network supporting short-distance communications among drones based on IEEE 802.11-based wireless local area network (WLAN) technologies. Therefore, the size of the network composed of one single mobile charging station and multiple drones is relatively small, and we assume that the flight time from drones' current positions to the mobile charging station is negligible. Thus, unexpected operational problems due to the delay induced by long flight time toward the mobile charging station are not considered in this paper.
Furthermore, we note that the specific design, system capabilities and state of the drones participating in the auction can vary in terms of battery capacity, residual battery, charging rates and so forth.
Formally, each drone $u_i$ is characterized by the battery capacity $c_i$, average amperage draw $e_i$, and battery residual charge $h_i$, which determines the mission lifetime. 
The average amperage draw $e_i$ denotes the amount of amperage required to the drone to operate on-board systems such as motors, embedded computers, sensors, etc. Each drone continuously monitors its own state and requests the scheduling of a charging slot to the mobile station if needed.

The requests from multiple drones to the mobile station for charging services can be interpreted as a distributed competition for a limited resource, which here is modeled and solved using an auction-based approach. In the considered setting, the auctioneer is the mobile station, which is also the owner and provider of the resource, and the drones are the buyers.

\begin{figure}
    \centering
    \includegraphics[width=0.8\columnwidth]{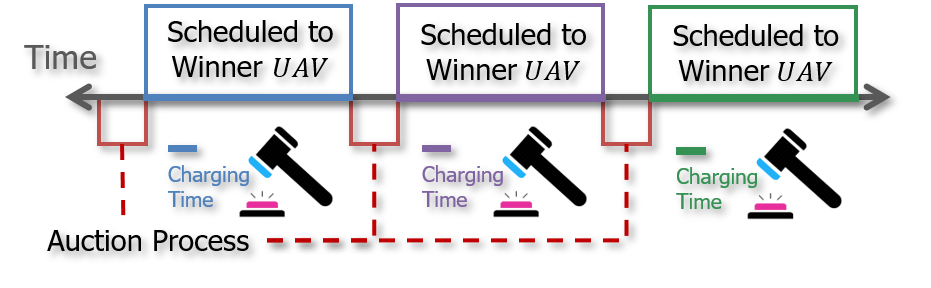}
    \caption{Auction procedure.}
    \label{fig:auctionmodel}
\end{figure}

The mobile station and drones exchange information, i.e., bids and other auction variables, over wireless links. The mobile station announces the start of the auction to the drones when the charging system is ready to serve (i.e., idle). Upon reception of the announcement, each drone makes its own private, and independent, valuation for the use of the charging system. The private valuation $v_i$ of drone $u_i$ is used to compete for the charging service. Note that the charging resource is assigned at the granularity of individual time slots as illustrated in Fig~\ref{fig:auctionmodel}. The mobile station $\mathcal{S}$ sells the charging service and obtains revenue $p_i$ paid by the winner drone $u_i$ via auction.

\BfPara{Drone Scheduling Auction Design} 
We use second price auction (SPA) as a baseline to design the auction in the considered setting. 
In SPA, all buyers submit their bids privately. The auctioneer receives the sealed bids and selects as winner the buyer who made the highest bid. The amount paid by the winner is set to be equal to the second highest bid value. 
Herein, the problem of assigning slots to drones is formulated as a single item auction based on SPA. Therefore, drones compete for one item, i.e., the charging service. Since the proposed approach is based on SPA, it is guaranteed that the charging service will be assigned to the drone with the highest valuation to the service~\cite{sujit2007distributed, lemaire2004distributed, bertuccelli2009real}.
Myerson presents provable analytical results for single item auctions optimizing the auctioneer revenue where each buyer has its own private valuation of the resource~\cite{myerson1981, dutting2017}. 

When the auction-based mechanism is designed, it is important to let the participants act truthfully to ensure system stability~\cite{myerson1981, dutting2017, luong2017, jiao2018, yen2018, khan2016}. Previous studies attempted to achieve this objective by enforcing truthfulness to individual participants. The concepts such as incentive compatibility (IC) and individual rationality (IR) are the characteristics of auctions inducing the truthful action of participants. Based on this approach, we use a Myerson auction where the following characteristic is used as the baseline mechanism:
The Myerson auction guarantees dominant strategy incentive compatibility (DSIC) and IR.
\theoremstyle{definition}
\begin{definition}(Incentive Compatibility ~\cite{bartal2003})
\label{def:IC}
    Incentive compatibility is defined by the following property: if for every bidder $j$, every valuation $v_j$, all declarations of the other bidders $v_{-j}$, and all possible "false declarations" $v'_j$, we have that bidder $j$'s utility with bidding $v'_{j}$ is no more than his utility with bidding the truth $v_j$. Formally, let $\lambda_j$ and $P_j$ be the mechanism 's output with input $(v_j, v_{-j})$ and $\lambda'_j$ and $P'_j$ be the mechanism's output with input $(v'_j,v_{-j})$, then $v_j(\lambda_j) - P_j > v_j(\lambda'_j) - P'_j$. 
\end{definition}

Thus, this weaker degree of DSIC guarantees IC, where IC means that the utility a participant can obtain by acting truthfully is greater than that by fake acting according to Definition~\ref{def:IC}.

\theoremstyle{definition}
\begin{definition}(Dominant Strategy Incentive Compatibility)
\label{def:DSIC}
    Dominant strategy incentive compatibility is defined as the following property. For each bidder $i$, and for every possible report of the other bidders bid $b_{-i}$, bidder $i$ weakly maximizes utility by reporting $b_i = v_i$. That is, for all possible reports $b^*_i$,
    $u_i(v_i, b_{-i}) \geq u_i(b^*_i, b_{-i})$.
\end{definition}

Thus, the DISC is a stronger degree of IC, meaning that a truthful action is a weakly dominant strategy, that is, the action is guaranteed to be the best, regardless of the actions of others, as shown in Definition~\ref{def:DSIC}.

\theoremstyle{definition}
\begin{definition}(Individual Rationality)
\label{def:IR}
    Individual rationality (IR) is defined by the following property: for every bidder $i$ and for every $v_i$, we have $v_i \geq p_i$, that is, no bidder is ever asked to pay more than its bid valuation.
\end{definition}

In a DSIC and IR auction, it is in the best interest of each bidder to report truthfully. Therefore, these characteristics make the overall auction truthful. The Myerson auction guarantees DSIC and IR, thus encouraging the bidders to report truthfully~\cite{myerson1981,dutting2017}.
Furthermore, the Myerson auction also guarantees auctioneer's revenue optimality. In the considered drone network model, we remark that the charging service controller obtains revenue by providing charging services. The following subsections describe in detail the components of the Myerson auction mechanism, i.e., private valuation, allocation rule, payment rule, reserve price, utility, and auction design.

\BfPara{Private Valuation} In the proposed auction, each drone $u_i$ has its own individual private valuation $v_i$. Each drone $u_i$ has an maximum battery capacity (denoted by $c_i$) and a current remaining battery (denoted by $r_i$). If the drone is assigned the mobile charging service time slot, the charged energy will be added to the residual energy in its own battery $r_i$. The amount of energy charged by the mobile charging station $S$ can be expressed as $q_i = \min(f\cdot t_i, (c_i - r_i))$ where $f$ is the charging rate per unit of time in the mobile charging station. The scheduled charging time to $u_i$ via auction is denoted by $t_i$. $t$ denotes the item being sold by auction, i.e., charging time. The higher $f\cdot t$, the higher the valuation of $t$ by drone $u_i$, and the drone is willing to pay a higher amount for the charging service. The expected drone flight time with current battery status is denoted as $l_i$, calculated as $l_i = \frac{r_i\cdot h_i}{e_i}$. If $l_i$ is larger, the drone will give a smaller valuation to $t$. Let $v_i$ denote the private valuation of drone $u_i$. Then, $v_i$ can be expressed as $v_i = \frac{f\cdot t}{l_i}$.

\BfPara{Allocation Rule} The allocation rule $g$ is used to determine the winner drone $u_i$ based on the valuation, i.e., to find which drone $u_i$ should be scheduled for charging. In the Myerson auction, the allocation rule that awards the item to the highest bidder is monotone. Therefore, in the proposed auction model, the allocation rule $g$ used to award the charging service to the highest bidder is monotone. Therefore, the allocation rule can be expressed as follows:

    \begin{equation}
        \label{eq:allocationrule_auction}
            u^* \in \arg \max_{u_i\in \boldsymbol u} g(v_i).
    \end{equation}

\BfPara{Payment Rule} The payment rule $p$ is used to determine the payment by the winner drone $u_i$ based on the valuation. In the proposed auction, the payment rule $p$ chooses a payment which is not higher than the private valuation, and it can be expressed as follows: 
    \begin{equation}
        \label{eq:paymentrule_auction}
            p_i(v_i) \in [0, x_i v_i]
    \end{equation}
where $x_i \in \left\{ 0,1 \right\}$, $\forall i = \left\{1,\dots,U\right\}$ stands for the variable to represent the winning valuation in the auction, and $u^*$ is the winner drone in the auction. 

\BfPara{Reserve Price} The proposed auction sets a specific price called a reserve price. The reserve price is the minimum reward the seller accepts~\cite{myerson1981}. In this paper, the reserve price is set to 0. In the auction, the auctioneer solicits the private bids from the bidders and computes the \textit{allocation rule} $g = (g_1, \dots, g_U)$ and \textit{payment rule} $p = (p_1,\dots, p_U)$.

\BfPara{Utility} The proposed auction guarantees DSIC and IR; and thus each bidder reports truthfully to maximize its own utility. Note that $x_i \in \left\{ 0,1 \right\}, \forall i = \left\{1,\dots,U\right\}$ stands for the variable corresponding to the winning in the auction. If the drone wins in auction, $x$ is set to $1$ whereas the $x$ is set to $0$ otherwise. Thus, the utility of drone $u_i$ can be calculated as $utility(u_i) = g(v_i) - x_i \cdot p_i(v_i), \forall i = \{1, \dots, U\}$.

\BfPara{Revenue Optimal Auction Design}
We define the \textit{virtual valuation} and \textit{virtual surplus} as in Myerson~\cite{myerson1981}. The virtual valuation $\phi_i(v_i)$ of a buyer $u_i$ in the auction is a function used to calculate the expected revenue of the auctioneer from that buyer $u_i$. The virtual surplus is the expected revenue excluding the computing cost defined below. In Myerson auctions, each bidder $i$ has its own individual private valuation $v_i$ which is drawn from the strictly increasing cumulative density function $F_i(v_i)$ where the probability density function of $v_i$ is denoted as $f_i(v_i)$~\cite{krishna2009auction, vickrey1961counterspeculation}. 
The virtual valuation of bidder $i$ with private valuation $v_i$ can be expressed as follows:
    \begin{equation}
        \label{eq:virvalue}
            \phi_i(v_i) = v_i - \frac{1-F_i(v_i)}{f_i(v_i)},
    \end{equation}

There is a cost in computing the outcome $c(g)$ which must be payed by the auction~\cite{hartline2006lectures}.
Given valuation $v_i$, virtual valuation $\phi_i(v_i)$, and allocation rule $g$, the virtual surplus can be calculated as follows:
    \begin{equation}
        \label{eq:virsurplus}
            \sum_{\forall i}\nolimits \phi_i(v_i)x_i - c(g).
    \end{equation}

In Myerson auction, the expected payment is proportional to the expected virtual surplus; and it can be computed as follows~\cite{myerson1981, hartline2006lectures}:
    \begin{equation}
        \label{eq:eqsurplus}
            \mathbb{E}_{b_i}[p_i(b_i)] = \mathbb{E}_{b_i}[\phi_i(b_i)x_i(b_i)].
    \end{equation}
    
    Therefore, if the virtual valuations $\phi(b)$ are non-decreasing in valuations $b$, the virtual surplus $\mathbb{E}_b$ is non-decreasing in valuations $b$. The bid $b$ is drawn from the distribution $F(b)$ with probability density function $f(b)$. Then, the expected payment can be computed as follows:
    \begin{eqnarray}
            \mathbb{E}_{b}[p(b)] &=&\int_{b=0}^h bg(b)f(b)\texttt{d}b-\int_{b=0}^h g(b)[1-F(b)]\texttt{d}b.\\
            &=&\int_{b=0}^h \left[b-\frac{1-F(b)}{f(b)}\right]g(b)f(b)\texttt{d}b.\\
            &=&\mathbb{E}_b[\phi(b)g(b)]. 
        \label{eq:expay}
    \end{eqnarray}

As a result, the proposed auction approach, which consists of a variant of the Myerson auction, is DSIC, IR, and revenue optimal.

However, Myerson auctions require full knowledge of the distributions $F_1, F_2, \dots, F_U$ according to Eq.~(\ref{eq:expay}). In the considered scenario, it is hard to obtain such information a priori, and we propose to use deep learning to estimate the distributions.  
In previous research results, it has been shown that the deep learning with limited structure can approximate specific functions~\cite{csaji2001, you2017deep, sill1998monotonic}. Specifically, herein, we use neural networks and unsupervised learning~\cite{dutting2017, luong2017} to approximate the virtual valuation function $\phi(v)$.
 The strength of deep learning is that the approximated function can be continually updated as inputs are acquired. The use of unsupervised learning makes the learning process possible, as it does not require the true values as input. The resulting auction is not only easily applicable to the distributed multi-drone network problem, but is also capable to adapt to continuously changing environments.

\begin{figure}
    \centering
    \includegraphics[width=0.9\columnwidth]{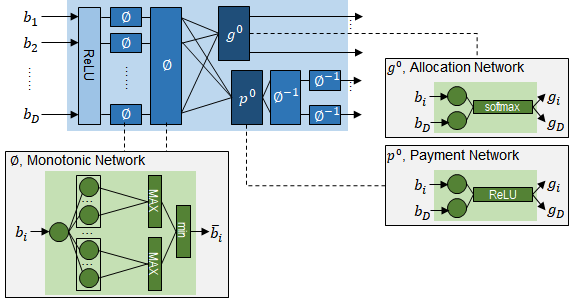}
    \caption{The proposed deep learning framework (revenue network) for revenue-optimal auction computation.}
    \label{fig:netsysmodel2}
\end{figure}

\section{Deep Learning based Auction Design}\label{sec:sec4_revision}
In this section, a deep learning based method for single item auctions is introduced. The method defines allocation rule $g$, payment rule $p$, and virtual valuation function $\phi$ for maximizing the revenue of the mobile charging station via deep learning. 
The deep learning model constitutes the auction that guarantees DSIC and IR as well as enables the revenue optimal computation for auctioneer~\cite{dutting2017,luong2017}. 
The revenue optimal auction can be configured through a relatively simple deep learning structure, i.e., composed of max/min operations and a loss function shaping the training process.

\begin{theorem}(Myerson~\cite{myerson1981}). 
\label{theo1}
There exist a collection of monotonically increasing functions $\phi_i$: $V_i \to R$, referred to as the virtual valuation functions, for selling a single item in the DSIC mechanism, which assigns the item to the buyer $i$ with the highest virtual value $\phi_i(v_i)$ assuming this quantity is positive and charges the winning bidder the smallest bid that ensures that the bidder is winning.
\end{theorem}

As mentioned earlier, the proposed deep learning based auction is a variant of Myerson auctions; and thus the bid set $\boldsymbol{b}$ is transformed to the virtual valuation $\overline{b_i}$ via virtual valuation transformation. Specifically, as expressed in Theorem \ref{theo1}, the bid set of $b_i, \forall i = \{1, \dots, U\}$ are converted to $\overline {b_i} = \phi^{mononet}(b_i), i = \{1, \dots, U\}$, where $\overline {b_i}$ denotes that the transformed bid of $u_i$. In this procedure, the trained deep network $\phi^{mononet}$ (a \textit{monotonic network}) is utilized to replace the virtual valuation function $\phi$. The $\phi^{mononet}$ consist of two layers, and is composed of linear computation units and min/max operation units.
Based on the transformed bid $\overline {b_i}$, the SPA with reserve price 0 (SPA-0) is performed. The SPA-0 calculates the allocation probability and the payment of the winner drone based on the rules (i.e., payment rule $p$ and allocation rule $g$) as follow:
\begin{theorem}(Myerson~\cite{myerson1981}).
\label{theorem2}
For any set of strictly monotonically increasing functions $\phi_1 , \dots ,  \phi_U : R_{\geq0} \to R_{\geq0}$, an auction defined by the allocation rule $g_i = \texttt{softmax}(\overline{b_i}) $ and payment rule $p_i = \phi^{-1}_i( \underset{j\neq i}{ \max } (\phi (b_i)) ) $ is DSIC and IR.
\end{theorem}

The $\phi^{mononet}$ should have non-decreasing monotone feature when converting $\boldsymbol{b}$ into transformed bid $\boldsymbol{\overline{b}}$. Therefore, the proposed deep learning network has a parameter constraint and a specific structure so that the deep learning network can be approximated to monotonic function via training process. The used parameters for deep learning, i.e., weights and biases, are positive. The structure of the network, shown in Fig~\ref{fig:netsysmodel2}, is rather simple. The two layers network $\phi^{mononet}$ is represented as $\phi^{shared}(\phi_i(b_i)), \forall i = \{1, \dots, U\}$~\cite{sill1998monotonic,dutting2017}.

The assignment rule consist of the \texttt{softmax} operation which has been used in deep learning based multimodal classification. The payment rules is composed of max operation and \texttt{ReLU}. 
The \texttt{ReLU} makes the transformed bid $\overline{b_i}$ which is less than the reserve price of SPA-0 to be 0. The max unit is used to make $\overline{p_i}$ be the highest transformed bid except $p_i$. The results of \texttt{ReLU} and max unit are denoted by $\overline{p}$. $\overline{p_i}$ is the value, before conversion to $p_i$, which should be paid by the winner drone $u_i$. Note that $\overline{p_i}$ can be larger than $b_i$. Therefore, in a IR auction, $\overline{p_i}$ can not be the payment. Thus $\overline{p_i}$ is converted to $p_i$ via $\phi^{-1}_{mononet}$. This process makes the result of deep learning based auction to be IR when revenue optimal auction is designed as shown in Fig~\ref{fig:netsysmodel2}. The $\phi^{-1}_{mononet}$ can be expressed as $\phi^{-1}_{shared}(\phi^{-1}_i(\overline{p_i})), \forall i=\{1,\dots,U\}$. The computations of $\phi^{-1}_{shared}$ and $\phi^{-1}_i$ is described as follows.

\begin{eqnarray}
    \label{revese:shared}
    p_i^\prime &=& \max_{1 \le g \le \mathcal{G}} \left\{ \min_{1 \le n \le N}(w^{shared}_{g,n})^{-1}(\overline{p_i} - \beta^{shared}_{g,n})  \right\}  \\
\label{reverse}
    p_i &=& \max_{1 \le g \le \mathcal{G}} \left\{ \min_{1 \le n \le N}(w^{i}_{g,n})^{-1}(p_i^\prime - \beta^{i}_{g,n})  \right\} 
\end{eqnarray}

The two layers network $\phi^{mononet}$ constitutes the virtual valuation function $\phi$ of Myerson auction, as shown in Fig~\ref{fig:netsysmodel2}. In the $\phi^{-1}_{mononet}$, it is important to reuse the weights from the $\phi^{mononet}$ network as presented in (\ref{reverse}) and (\ref{revese:shared}). This forces $b_i$ to be equal to $\phi^{-1}_{mononet}(\phi^{mononet}(b_i))$ as in the case in which the Myerson virtual valuation function is based on full knowledge of the distribution $F$. 
The result of $\phi^{-1}_{mononet}$ is the payment which should be paid by winner drone $u_i$. Therefore, the result $p_i$ is greater than the second highest bid of $\boldsymbol{b}$ and smaller than the winning bid $b_i$~\cite{myerson1981}.

Additional networks are required to implement the rules in overall auction processes. 
The deep learning networks used in the proposed auction consist of three modular networks as follows:
    (i) a network that can replace the virtual valuation function $\phi_i$ of the Myerson auction,
    (ii) a network for the allocation rule $g_i$,
    and 
    (iii) a network for the payment rule $p_i$.
The above networks are optimized according to a loss function via a training process. 
The loss function is essential to enable the deep learning computation of the same structure to have different characteristics~\cite{arjovsky2017wasserstein, mao2017least}, and plays an important role in deep learning.

\BfPara{Loss Function}
In this paper, the negative expected virtual surplus is used as a loss function where the virtual surplus is equivalent to the revenue of the mobile charging station, that is, the auctioneer and seller. The loss function is used to train deep neural network parameters (weights and biases). The deep neural network that configures a revenue-optimal auction is composed of weights (denoted as $\boldsymbol{w}$) and biases (denoted as $\boldsymbol{\beta}$) which replaces the virtual valuation function of Myerson.  
Hhere, the deep neural network model automatically learns the distribution, and fits its parameters to the actual distribution of the data during the training process. 
The trained neural networks are approximated by a virtual valuation function which is based on fully distributed knowledge. The parameters $\boldsymbol{w}$ and $\boldsymbol{\beta}$ of the deep neural network are trained through unsupervised learning without ground truth information, i.e., the winner (which drone will be scheduled for charging) and payment (how many the winner drone will pay). Therefore, the results of the allocation and payment rules are used for training parameters (i.e., $\vec{w}$ and $\vec{\beta}$) can be explained as follows:
\begin{equation}
\label{Eq8}
\mathcal{R}\left(\vec{w},\vec{\beta}\right) = - \sum^{U}_{i=1}\nolimits g_i(b_i) \ast p_i(b_i)
\end{equation}
where the loss function \eqref{Eq8} stands for the expected negative revenue of auctioneer, i.e., the maximization of the expected revenue of the auctioneer since the loss function should be minimized eventually during the training procedure. 
Based on the loss function, the benefit of the deep learning based auction is seen in the training process. The proposed networks which replace the virtual valuation function as well as auction rules are optimized to DSIC, IR and the revenue optimal auction.

\BfPara{Deep Learning Training} 
The detailed training process of the three networks are summarized in Algorithm~\ref{algo:three}, where, based on the bid $\boldsymbol{b}$, the payment and allocation probabilities are calculated (line $[4-9]$). $\mathcal{R}(w,\beta)$ is the loss function to guide the deep learning network training. The negative expected revenue is used as the loss function. The loss function can be calculated by allocation probability $g$ and payment $p$ (line $[10]$). The $L_2(w,\beta)$ is regularization factor which are used to regularize the deep learning parameters (weights and bias) (line $[11]$). The $L_2(w,\beta)$ regularization prevents parameters from becoming excessively large. The training process is based on unsupervised learning; and thus the allocation probability and payment are the only required information. This means that the environmental information such as distribution of private valuation is not required. As a result, the proposed deep learning network can be easily applied to mobile charging stations. The parameters are determined by means of empirical experiments as the payments of winners are updated sensitively due to the weight range (line $[15-16]$).

\subsection{Deep Learning Networks}
In this paper, the virtual valuation function is replaced by the two layers network $\phi^{mononet}$, composed of the monotonic networks.

\BfPara{Monotonic Network}
As shown in Fig~\ref{fig:netsysmodel2}, the monotonic network is a three-layer deep neural network. The input layer is configured with multiple groups composed of sets of linear units. The maximum value of each group is calculated in the second layer. The last layer selects the minimum value of the given output of the second layer. As the name suggests, the monotonic network is monotonic, and this characteristic is preserved regardless of the number of groups, units, and the order of min/max operations.

\setlength{\textfloatsep}{0pt}
\begin{algorithm}[t!]
\SetInd{0.5em}{0.8em}
\small
\setstretch{1.0}
    \SetKwInOut{Input}{Input}
    \SetKwInOut{Output}{Output}
    \SetKwInOut{Initialize}{Initialize}
    \Input{ $k, U, \boldsymbol{\mathcal{B}}=\left\{\boldsymbol{\mathcal{B}_1, \dots, \mathcal{B}_\mathcal{T}}\right\}$ where each input set $\boldsymbol{\mathcal{B}_t}\triangleq(b_1, \dots , b_{U}) $}
    \Output{ Optimized weights $\boldsymbol{w}$ and $\boldsymbol{\beta}$}
    \Initialize{ The network weights $\boldsymbol{w}$ and $\boldsymbol{\beta}$ using Xavier initialization}
    \While{epoch r: $1 \to \mathcal{R}$}{
        \While{ $\boldsymbol{t}: 1 \to \mathcal{T}$}{
              {\bf{Forward:}}\\
              $\triangleright$ $b_i^\prime = \phi_i{(b_i)} = \underset{1 \le g \le \mathcal{G}}{\min}\left\{ \underset{1 \le n \le \mathcal{N}}{\max}(w^{i}_{g,n}b_{i}+\beta^{i}_{g,n})  \right\}$\; 
              $\triangleright$ $\overline{b_i} = \phi_{shared}(b_i^\prime)$\;
              $\triangleright$ $g_i = \texttt{softmax}\left(\overline{b_1}, \dots, \overline{b_U};k\right) = \frac{e^{k\overline{b_i}}}{\sum_{j=1}^{U} e^{k\overline{b_j}}}$\;
              $\triangleright$ $\overline{p_i} = \texttt{ReLU}\left\{ \max_{j\neq i} ( \overline{b_i}) \right\}$\;
              $\triangleright$ $p_i^\prime$ = $\phi_{shared}^{-1}(\overline{p_i})$\;
              $\triangleright$ $p_i = \phi^{-1}_i(p_i^\prime) = \underset{1 \le g \le \mathcal{G}}{\max} \left\{ \underset{1 \le n \le \mathcal{N}}{\min}(w^{i}_{g,n})^{-1}(p_i^\prime - \beta^{i}_{g,n})  \right\} $\;
              $\triangleright$ Compute the expected negative revenue          $\mathcal{R}(w,\beta) = - \sum^{\mathcal{U}}_{i=1} g_i(b_i) * p_i(b_i)$\;
              $\triangleright$ Compute $L_2$ weight loss
              $L_2(w, \beta) = \sum^{U}_{i=1}\sum^{\mathcal{G}}_{g=1} \sum^{\mathcal{N}}_{n=1}\left\{(w^i_{g,n})^2 + (\beta^i_{g,n})^2\right\}$\;
              $\triangleright$ Compute 
              $\texttt{Loss}(w, \beta)$ = $\mathcal{R}(w,\beta) + L_2(w, \beta)$ 
              \\
              {\bf{Optimize:}}\\
              $\triangleright$ Update $\vec{w}$ and $\vec{\beta}$ for minimizing $\texttt{Cost}(w, \beta)$\;
              $\triangleright$ Clip $\vec{w}$ $(\min B, \max \infty)$\;
              $\triangleright$ Clip $\vec{\beta}$ $(\min 0, \max \infty)$\;
        }
    }
\caption{Deep Learning Training}
\label{algo:three}
\end{algorithm}

\BfPara{Virtual Valuation Network}
The virtual valuation function in the Myerson auction is replaced with the monotonic network. The computation of $\phi^{shared}$ and $\phi_i$ is implemented as follows.
\begin{equation}
\small
\label{Eq3:shared}
    \overline{b_i} = \phi_i^{shared}{(b_i^\prime)} = \underset{1 \le g \le \mathcal{G}}{\min} \left\{ \underset{1 \le n \le \mathcal{N}}{\max}(w^{shared}_{g,n}b_{i}^\prime + \beta^{shared}_{g,n})  \right\} 
\end{equation}
\begin{equation}
\label{Eq3}
    b_i^\prime = \phi_i{(b_i)} = \underset{1 \le g \le \mathcal{G}}{\min} \left\{ \underset{1 \le n \le \mathcal{N}}{\max}(w^{i}_{g,n}b_{i} + \beta^{i}_{g,n})  \right\} 
\end{equation}

The bid $b_i$ of the drone $u_i$ is transformed to $\overline{b_i}$ via the virtual valuation network $\phi^{mononet}$. In the $\phi^{mononet}$, all outcomes of $\phi^{shared}$ are calculated on the same weights, whereas the $\phi_i$ calculates the outcome using different weights for each bid. The inverse computation of $\phi^{mononet}$ is denoted by $\phi^{-1}_{mononet}$. The $\phi^{-1}_{mononet}$ that determines the payment of the winner drone $u_i$ which is composed of two networks. In the computation of $\phi^{-1}_{mononet}$, the weights of $\phi^{mononet}$ are used. Thus the computations of two layers can be expressed as follows:
\begin{equation}
\begin{medsize}
\label{Eq4:shared}
    p_i^\prime = \phi^{-1}_{shared}(\overline{p_i}) = \underset{1 \le g \le \mathcal{G}}{\max} \left\{ \underset{1 \le n \le \mathcal{N}}{\min}(w^{shared}_{g,n})^{-1}(\overline{p_i} - \beta^{shared}_{g,n})  \right\}
\end{medsize}
\end{equation}
\begin{equation}
\label{Eq4}
    p_i = \phi^{-1}_i(p_i^\prime) = \underset{1 \le g \le \mathcal{G}}{\max} \left\{ \underset{1 \le n \le \mathcal{N}}{\min}(w^{i}_{g,n})^{-1}(p_i^\prime - \beta^{i}_{g,n})  \right\}
\end{equation}

The payment $\overline{p_i}$ of the drone $u_i$ is transformed to $p_i$ via the $\phi^{-1}_{mononet}$. The $\phi^{-1}_{mononet}$ consists of $\phi^{-1}_i$ and $\phi^{-1}_{shared}$. The same weights are used to calculate all outcomes of $\phi^{-1}_{shared}$. The outcome of $\phi^{-1}_i$ is calculated based on different weights for each bid, as shown in (\ref{Eq4}).

The monotonic network is responsible for the transformation of the virtual bid $\overline{b_i}$ in auction. As mentioned above, the optimal revenue is equivalent to the optimal virtual surplus. Thus, the monotonic network is major component of the auction. However, in order to configure the revenue-optimal auction, the additional network by allocation and payment rules is required. In this paper, the payment and allocation rules are configured with \texttt{ReLU} and \texttt{softmax} which have been mainly used in deep learning as an activation function. This makes backpropagation easy during the training process.

\BfPara{Allocation Rule Network ($g_i$)}
This section describes in more details the structure of the allocation rules ($g_i$). In this paper, since this allocation rule is implemented using a deep neural network, the probability is calculated using \texttt{softmax} which converts the input vector into a probability vector. 
The allocation rule $g$ awards the charging service to the highest bidder drone; and thus the highest probability is assigned to the highest bidder.
The continuous function (\ref{eq:paymentrule_auction}) traditional auction is approximated using the deep network, which converts the input vector $\overline{b}$ to the probability vector. 
In the SPA auction with reserve price 0 (SPA-0), the allocation rule assigns the highest winning probability to the highest bidder whose transformed bid $\overline{b_i}$ is greater than $0$, $\overline{b_i}>0$. 
The \texttt{softmax} based assignment can be calculated as follows: 
\begin{eqnarray}
\label{Eq5}
g_i = \texttt{softmax}\left(\overline{b_1}, \dots, \overline{b_U};k\right) = \frac{e^{k\overline{b_i}}}{\sum_{j=1}^{U} e^{k\overline{b_j}}}.
\end{eqnarray}
The parameter $k$ is a constant value and it determines the quality of the approximation. As the $k$ increases, the quality of the approximation increases, whereas the smoothness in the allocation network decreases. For simplicity, this means that the higher $k$ makes a large difference between the allocation probabilities of users~\cite{dutting2017}. When the networks are trained to minimize (\ref{Eq8}), the value of  $g_i(b_i)$ increases. As a result, since the profit of the auctioneer is related to the second highest $g_i(b_i)$, it is also a function of the parameter $k$.
Results in Sec.~\ref{sec:sec4}) show how larger values of $k$ lead to higher profits.

\BfPara{Payment Rule Network ($\overline{p_i}$)}
This section describes the structure of the allocation rule ($\overline{p_i}$). 
The \texttt{ReLU} is widely used in deep learning computation as an activation function. In the proposed auction, the payment $p_i$ of drone $u_i$ is calculated from the transformed bid $\overline{b_i}$. Before the computation of $\phi^{-1}_{mononet}$, the deep network excludes the bid below the reserve price 0 via \texttt{ReLU}$(\overline{b_i})$ $\triangleq \max(\overline{b_i}, 0)$. The input $\overline{b_i}$ is the second highest transformed bid which is the output of $\max_{j\neq i}\left( \overline{b_i}\right)$. The payment rule network can be, then, calculated as:
\begin{equation}
\label{Eq6}
\overline{p_i} = \texttt{ReLU} \left\{\max_{j\neq i}\left( \overline{b_i}\right) \right\}
\end{equation}
and the result $\overline{p_i}$ is used as an input of (\ref{revese:shared}), i.e., the actual payment of winner drone.

\setlength{\textfloatsep}{0pt}
\SetInd{0.5em}{0.8em}
\begin{algorithm}[t]
\small
\setstretch{1.00}
    \SetKwInOut{Input}{Input}
    \SetKwInOut{Output}{Output}
    \Input{$t$, $f$, Bid sets $\boldsymbol{b}\triangleq (b_1, \dots , b_U)$ }
    \Output{allocation probability set $\boldsymbol{g_i} \triangleq (g_1, \dots, g_U)$,\\ payment set $\boldsymbol{p_i} \triangleq (p_1,\dots,p_U)$}
    \While{Mobile charging system is $\boldsymbol{idle}$}{
        $\triangleright$ Drones: charging scheduling valuation $v_i$\;
        $\triangleright$ Drones: submit bid $b_i$\;
        $\triangleright$ $b_i^\prime = \phi_i{(b_i)} = \underset{1 \le g \le \mathcal{G}}{\min}\left\{ \underset{1 \le n \le \mathcal{N}}{\max}(w^{i}_{g,n}b_{i}+\beta^{i}_{g,n}) \right\}$\; 
        $\triangleright$ $\overline{b_i} = \phi_{shared}(b_i^\prime)$\;
        $\triangleright$ $g_i = \texttt{softmax}\left(\overline{b_1}, \dots, \overline{b_U};k\right) = \frac{e^{k\overline{b_i}}}{\sum_{j=1}^{U} e^{k\overline{b_j}}}$\;
        $\triangleright$ $\overline{p_i} = \texttt{ReLU}\left\{ \max_{j\neq i} ( \overline{b_i}) \right\}$\;
        $\triangleright$ $p_i^\prime$ = $\phi_{shared}^{-1}(\overline{p_i})$\;
        $\triangleright$ $p_i = \phi^{-1}_i(p_i^\prime) = \underset{1 \le g \le \mathcal{G}}{\max} \left\{ \underset{1 \le n \le \mathcal{N}}{\min}(w^{i}_{g,n})^{-1}(p_i^\prime - \beta^{i}_{g,n})  \right\} $\;
        $\triangleright$ Calculate winner and payment $(\boldsymbol{g_k, p_k})$\;
        $\triangleright$ Winner Drone: Pay payment\;
        $\triangleright$ Allocate charging system to the winner\;
    }
\caption{Deep Learning-Based Algorithm for the Auction Controlling the Charging Scheduling}
\label{algo:auction}
\end{algorithm}

\subsection{Overall Auction Mechanism}
The overall deep learning-based auction mechanism is summarized in Algorithm~\ref{algo:auction}. If the mobile charging system becomes idle, the auction is initiated (line $[1]$). The valuation $v_i$ for the charging time is computed by each drone $u_i$ based on its own private criteria. Then, based on the individual private valuation, each drone submits its bid $b_i$  (line $[2-3]$). The mobile charging station runs the auction using the pre-trained networks. If 
$\boldsymbol{\overline{p}} = \boldsymbol{0}$, then all the drones assign a low valuation to the charging time and the mobile charging system does not allocate the charging time to users. If there exist bids which are larger than reserve price 0, the corresponding allocation and payment probabilities are calculated using the proposed deep learning networks, i.e., virtual valuation network, allocation network, and payment network (line $[4-10]$).
Because the proposed deep learning auction is the variant of SPA-0, any bid below the reserve price 0 is converted to 0 (line $[7]$).
As shown in line $[11]$, the mobile charging station assigns the payment of $p_i$ to the drone $u_i$ with the highest $g_i$. Finally, the mobile charging station allocates the charging time to the winner drone $u_i$ (line $[12]$). The drone, then, reaches the charging station and occupy it for the duration of the slot. After the winner drone leaves the charging station, next iteration starts if the mobile station is idle.

\begin{figure*}[t!]
\centering
  \begin{subfigure}{0.45\textwidth}
            \centering
            \includegraphics[width=1\textwidth]{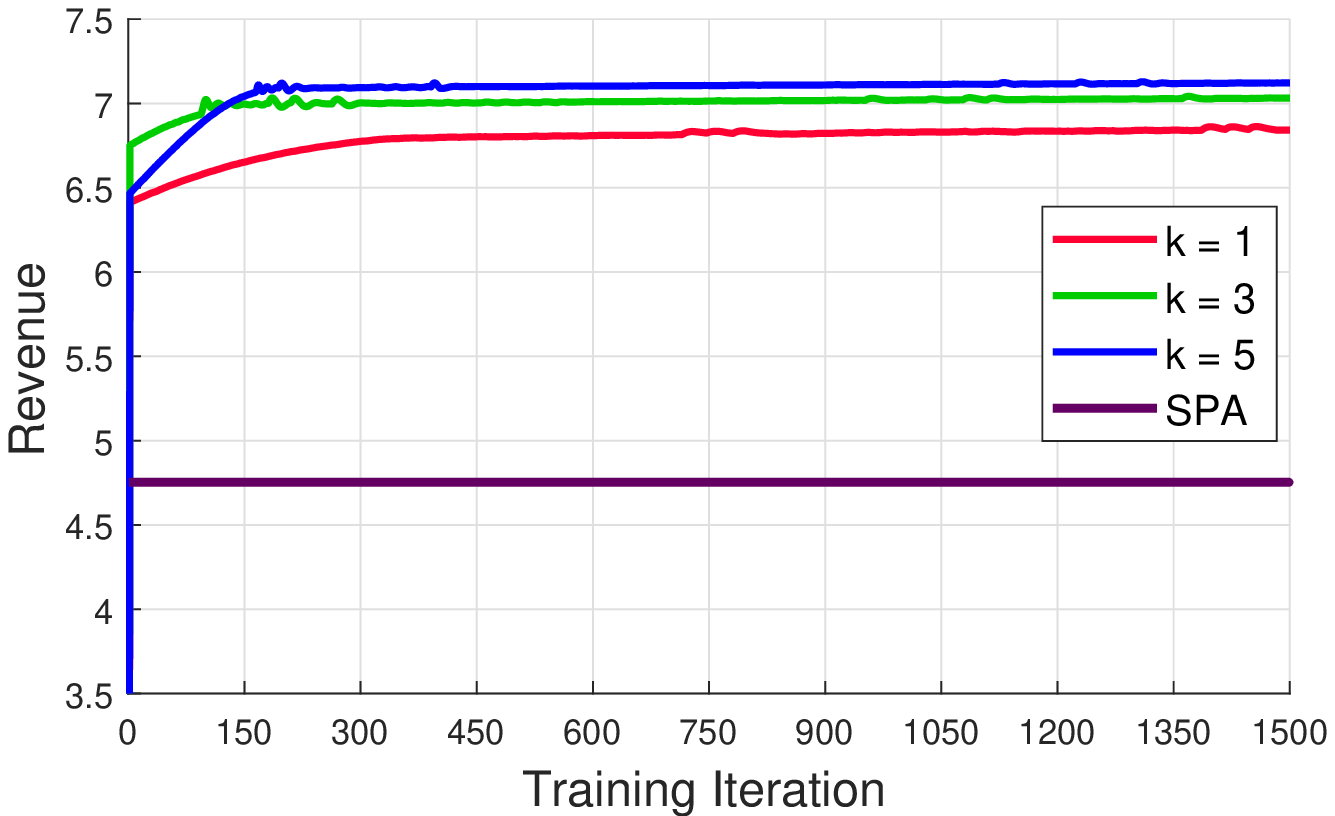}
            \caption{Revenue statistics, 5 drones}
            \label{fig4:payU5}
  \end{subfigure}
  \begin{subfigure}{0.45\textwidth}
            \centering
            \includegraphics[width=1\textwidth]{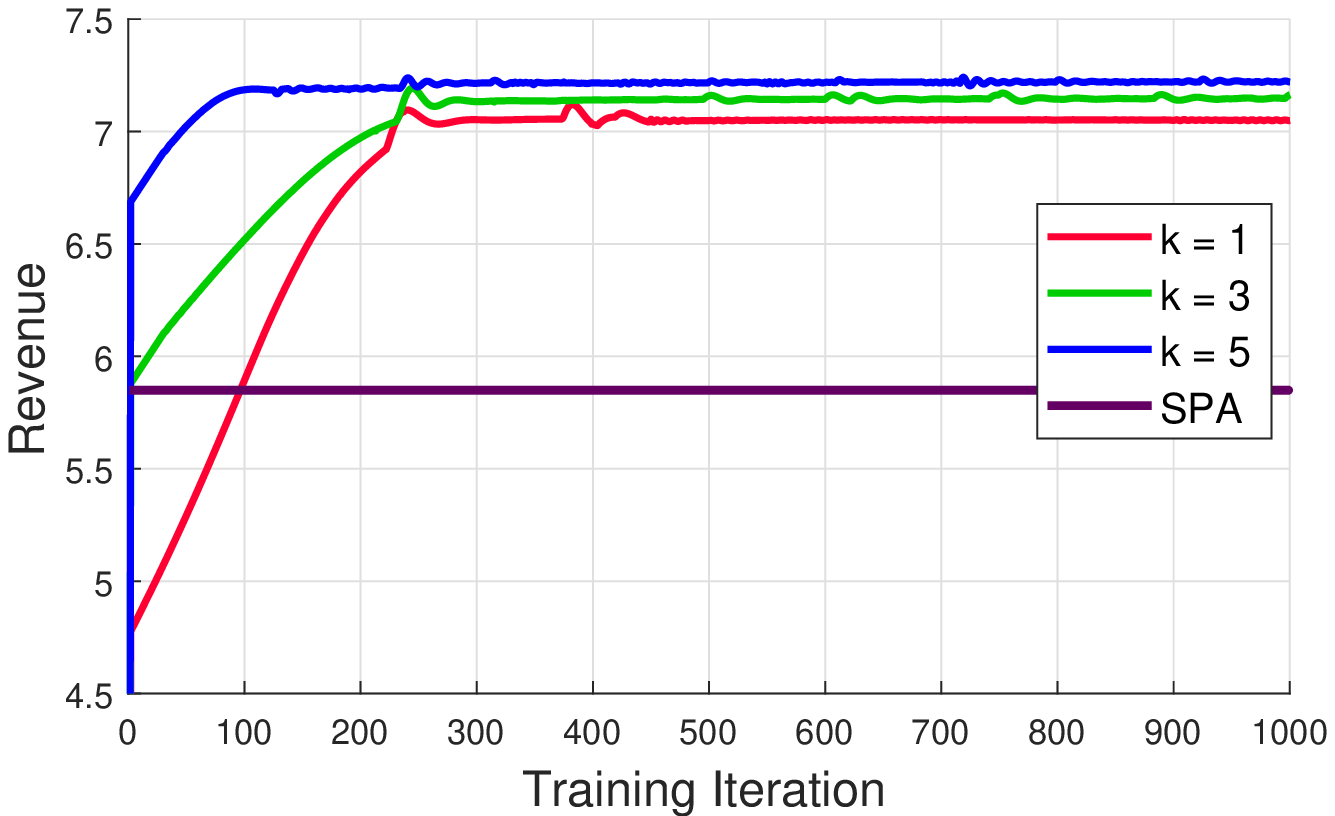}
            \caption{Revenue statistics, 10 drones}
            \label{fig4:payU10}
  \end{subfigure}
  \begin{subfigure}{0.45\textwidth}
            \centering
            \includegraphics[width=1\textwidth]{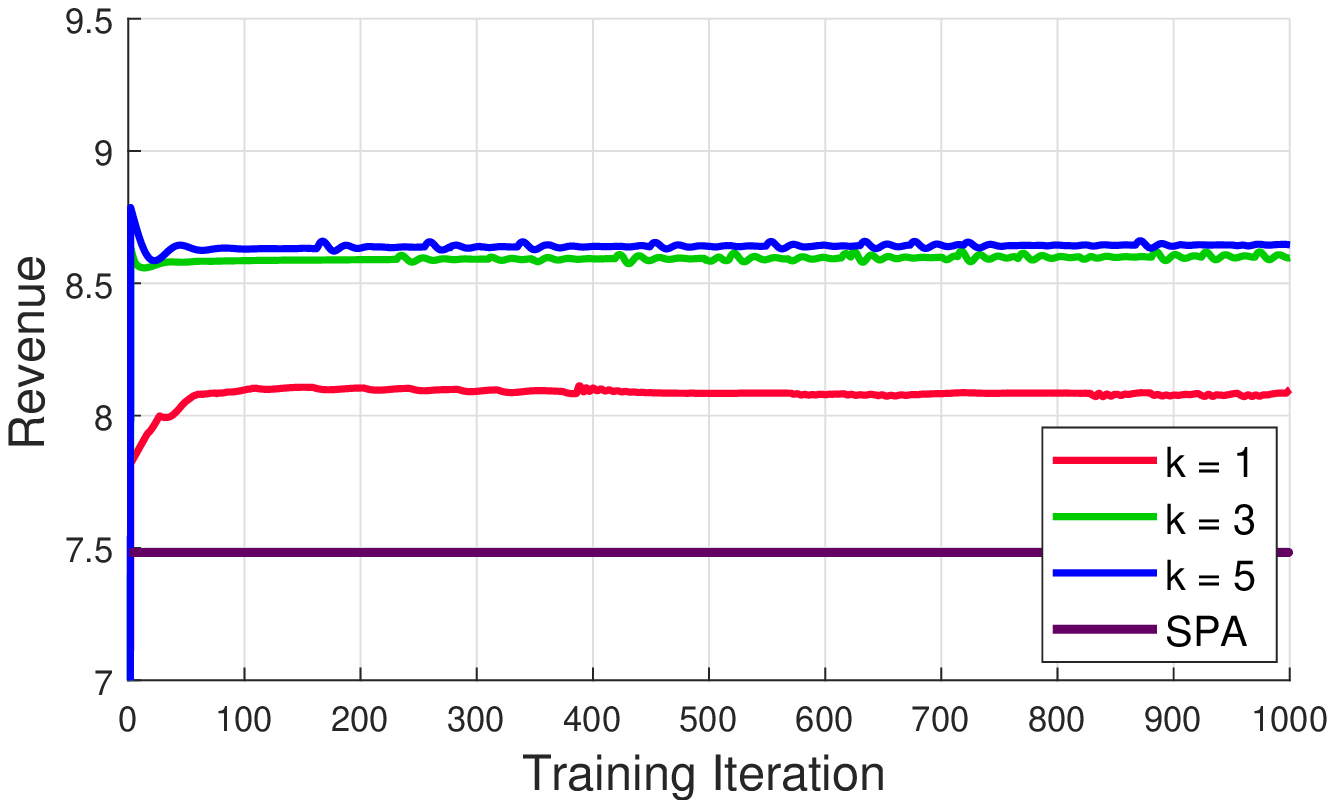}
            \caption{Revenue statistics, 15 drones}
            \label{fig4:payU15}
  \end{subfigure}
  \begin{subfigure}{0.45\textwidth}
            \centering
            \includegraphics[width=1\textwidth]{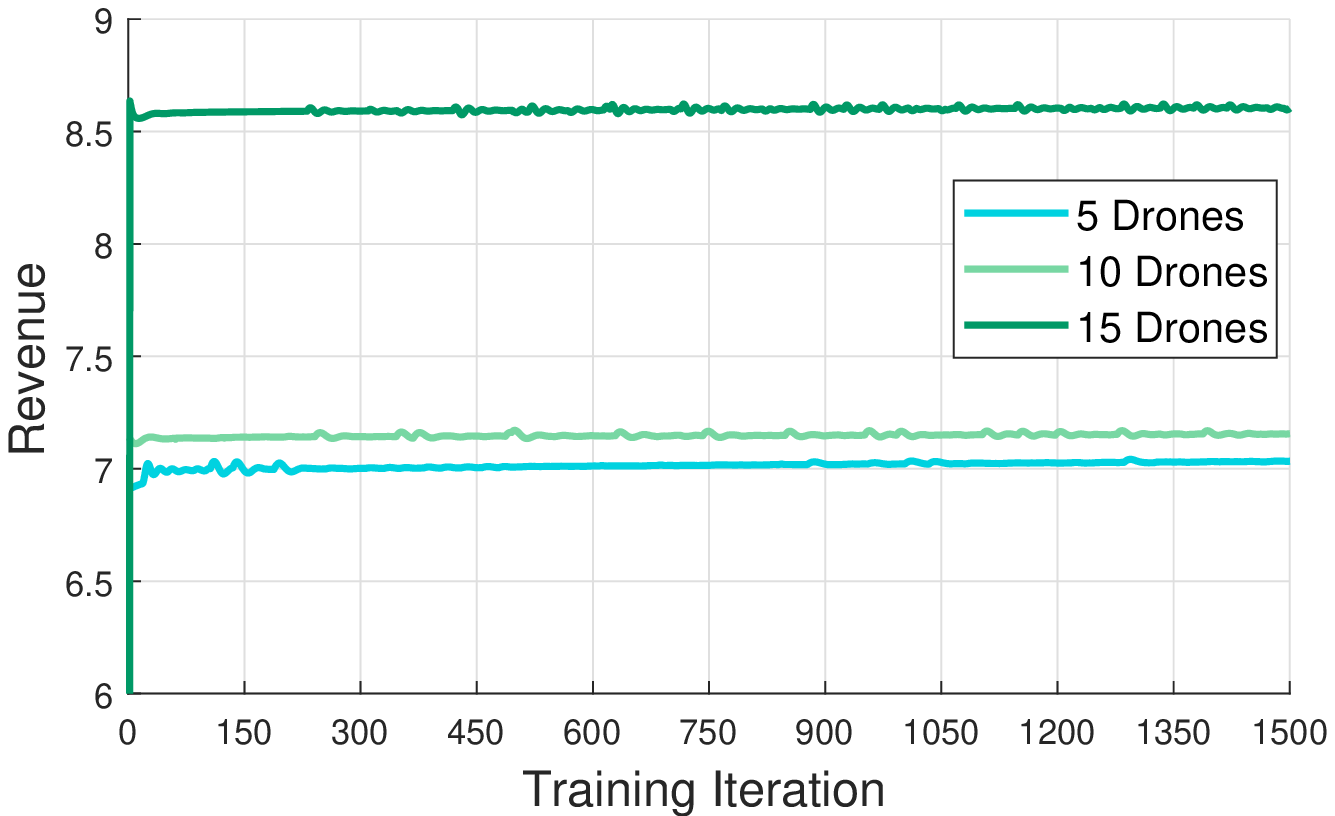}
            \caption{Revenue statistics, 5/10/15 drones, $k = 3$}
            \label{fig4:payU51015}
  \end{subfigure}
  \caption{Revenue changes by $k$ and $U$.}
  \label{fig:fig4}
\end{figure*}

\begin{table}
\footnotesize
        \caption{Revenue changes by $k$, $U$ (in Fig.\ref{fig4:payU5}-\ref{fig4:payU15})}
        \label{tab:self}
        \begin{center}
            \scalebox{1}{
            	\begin{tabular}{l|c|c|c|c}
                    \toprule
                      & SPA & $k = 1$ & $k = 3$ & $k = 5$\\
                    \midrule [1.0pt] 
                    5 drones & 4.7532 & 7.0001  &  7.0121 & 7.1009 \\
                    10 drones & 5.8493  & 7.0345 & 7.1408 & 7.2235 \\
                    15 drones & 7.4829 & 8.0912 & 8.6038 & 8.6471 \\[0.2ex]
                    \bottomrule
            	\end{tabular}
            	}
        \label{tab:fig4}
        \end{center}
\end{table}

\begin{table}
\footnotesize
\caption{Parameters}
\label{tab:parameters}
\begin{center}
    \scalebox{1}{
	\begin{tabular}{l|c}
    \toprule
    Variables & Descriptions \\
    \midrule [1.0pt]
    The number of drones& 5, 10, 15\\ 
    Learning rate & 0.0001\\
    $L_2$ regularization parameter & 0.001\\
    Training set size & 100000 bid sets\\
    Simulation epoch & 100\\
    Approximate quality $k$ & 1, 3, 5\\
    Distribution of $l_i$ & U[1:5], U[5:10], U[1:10]\\
    Weight range B & 0.0001\\[0.2ex] 
    \bottomrule
	\end{tabular}
	}
\label{tab:hresult}
\end{center}
\end{table}

\section{Performance Evaluation}\label{sec:sec4}

\BfPara{Software Prototype}
First, we describe the software developed to test the auction mechanism.
The \textit{Xavier initializer} was used for weight value initialization, where the biases were initialized as 0. As mentioned earlier, $L_2$ regularization is used to prevent excessive parameter growth during training and reduce overfitting. 
The regularization factor was set to $0.001$. During the training phase, the \text{Adam}~\cite{kingma2014adam} optimizer was us to iteratively. This choice is motivated by the need to keep separated the learning rates for each weight. An \textit{exponentially decaying average} of previous gradients was used for iteration-based optimization. In the experiments, different uniform distributions were used for data generation, as shown in Table \ref{tab:parameters}. Data-intensive evaluation was conducted with $100,000$ generated data sets. Among the data sets, $70$\% of sets were used for training; and the remaining $30$\% were used for testing. The proposed deep learning-based auction mechanism was implemented in Python/TensorFlow~\cite{tensorflow} and Keras~\cite{keras}. 
A multi-GPU platform (equipped with 2 NVIDIA Titan XP GPUs using 1405\,MHz main clock and 12\,GB memory) was used for training and testing.

\BfPara{Experimental Setting}
The test environment includes 5, 10, or 15 drones.
During performance evaluation, the parameter $k$ is determined to control the quality of approximation. First, we compare the proposed model with SPA-0 with a priori knowledge to demonstrate revenue-optimality.  Results show the ability of the proposed deep learning-based approach to adapt to different scenarios. The valuation results of drones are generated based on various distributions as defined in Table~\ref{tab:parameters}. Table~\ref{tab:hresult} summarizes the used parameters. 

\begin{figure*}[t]
  \centering
  \begin{subfigure}{.48\textwidth}
            \includegraphics[width=1\textwidth]{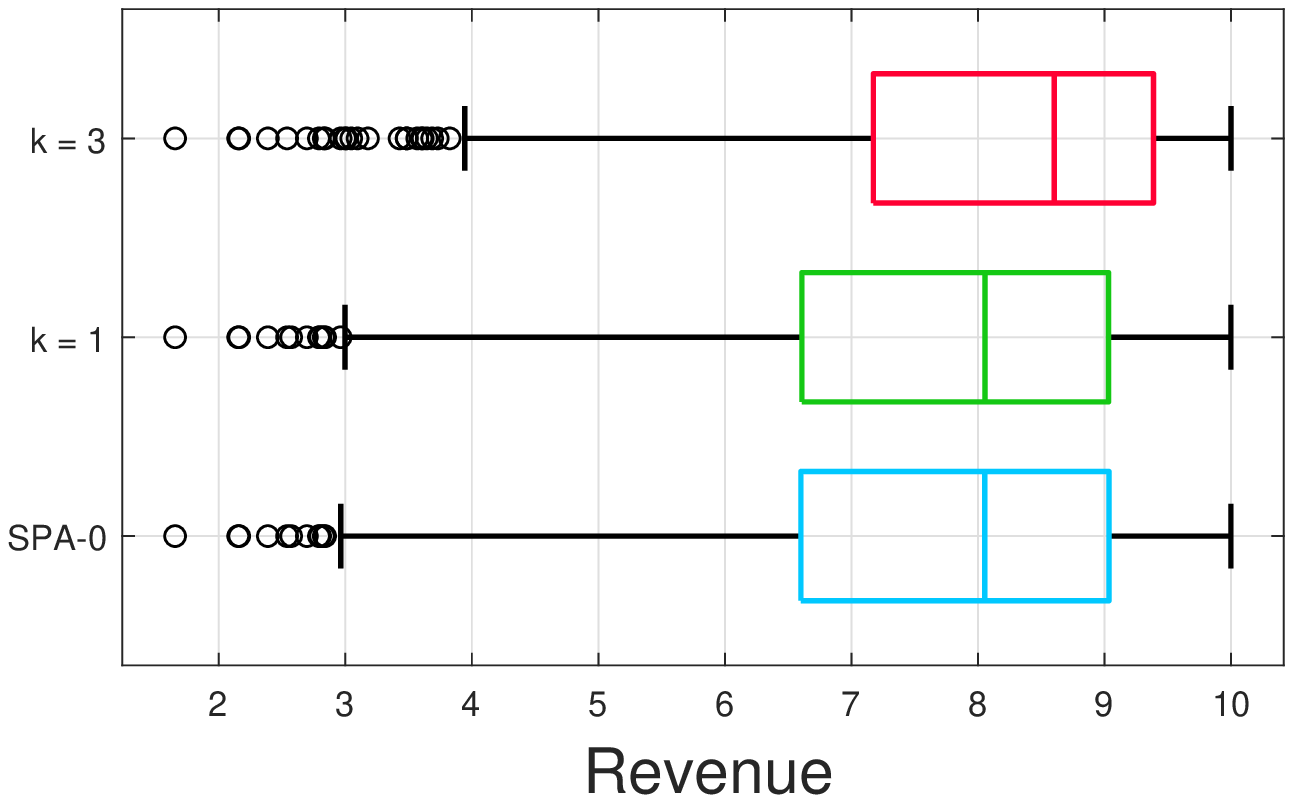}
            \caption{Revenue statistics of mobile charging station.}
            \label{fig7:box}
  \end{subfigure}
   \begin{subfigure}{.48\textwidth}
            \includegraphics[width=1\textwidth]{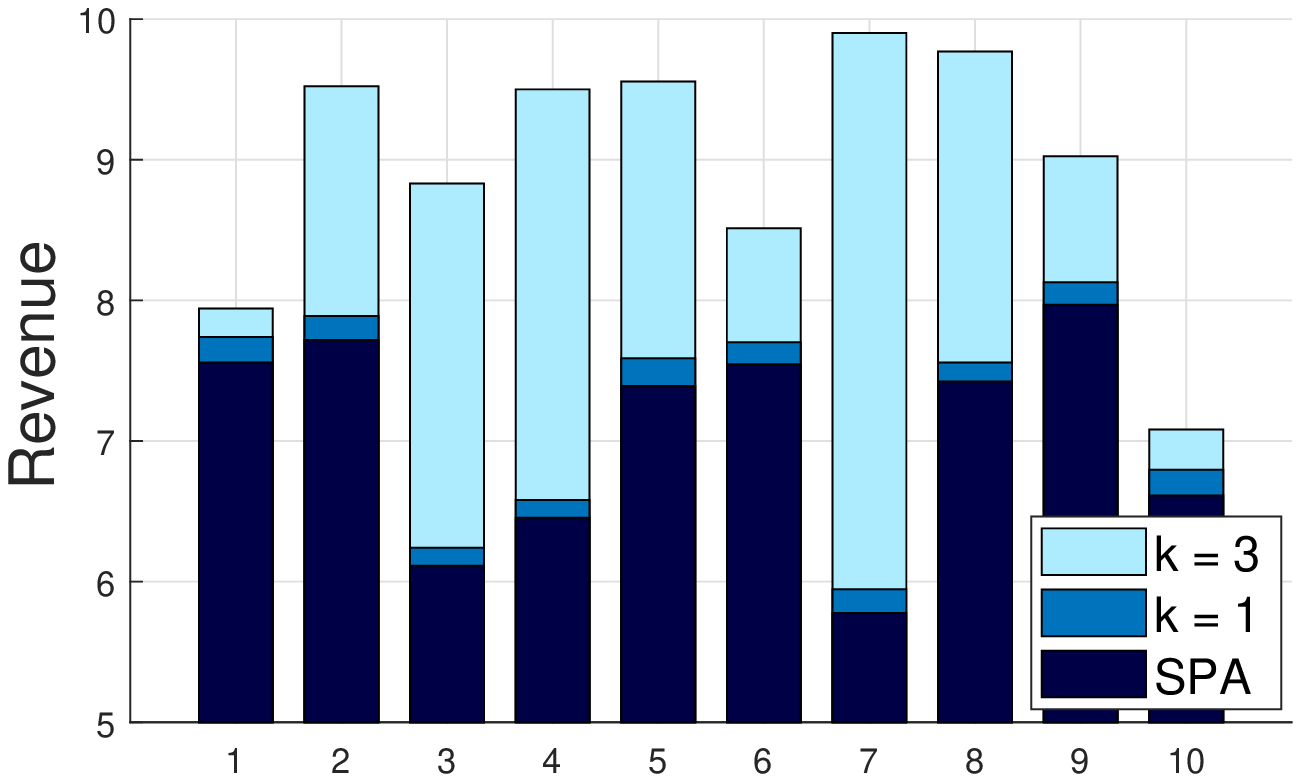}
            \caption{Revenue of mobile charging station.}
            \label{fig7:bar}
  \end{subfigure}
  \caption{Revenue analysis.}
  \label{fig:fig7}
\end{figure*}

\BfPara{Revenue Analysis - Parameter $k$}
The proposed framework is based on the Myerson optimal auction, which produces an increased revenue to the mobile charging station compared to SPA-0 auctions. The experiments shown in Fig.~\ref{fig:fig4} confirm this effect, and illustrate the effect of the parameter $k$.
Fig.~(\ref{fig4:payU5})-(\ref{fig4:payU15}) show a comparison between the revenue of the mobile charging station -- the auctioneer -- as a function of the parameter $k$ defined in \eqref{Eq5}. In the experiments, the bid set is uniformly generated in the range of $0-10$. The bid is calculated based on the private valuation as discussed in Sec.~\ref{sec:sec3}. The value of system parameters, such as battery consumption rate, and weight, is also assumed to be uniformly distributed. The results in Fig.(\ref{fig4:payU5})-(\ref{fig4:payU15}) show that the revenue increases as the $k$ increases. The numerical results are presented in the Table~\ref{tab:fig4}. The revenue gap between the SPA-0 and the proposed auction when the number of drone is $5$ is near $2.25$ when $k=1$, near $2.26$ when $k=3$ and about $2.35$ when $k=5$. The mobile charging station can take the highest revenue, i.e., the case where $k=5$. This result shows that the revenue of mobile charging station increases in the order of $SPA-0$, $k=1$, $k=3$ and $k=5$. The parameter $k$  determines not only the approximation quality of \texttt{softmax} function but also the revenue of charging station.
In Fig.~\ref{fig:fig4}, the number of drones which participate in the proposed charging scheduling auction is updated. In general, more drones participate in auction, the higher the bid can be submitted to the auction with high probability; and thus the second highest bid value of the auction can be increased while the number of drones increases. Note that the revenue of auctioneer is equivalent to the payment of user. Therefore, the payment of winner drone increases. As a result, the revenue of mobile charging station becomes larger. In the SPA-0, the revenue is increased from $4.7532$ to $5.8493$ when the number of drones increases from $5$ to $10$. Similarly, the revenue of proposed charging scheduling auction increases to $7.0345$ when $k=1$, $7.1408$ when $k=3$ and $7.2235$ when $k=5$. This tendency is maintained when the number of drones increases from $10$ to $15$, as shown in (Fig.\ref{fig4:payU5}-\ref{fig4:payU15}). Fig.~\ref{fig4:payU51015} and Table.~\ref{tab:fig4} shows the revenue of $k=3$ model when the number of drones increases from $5$ to $15$. In this evaluation, the training of deep learning network uses the pre-trained weights. Based on this experiment result, we can confirm that the proposed deep learning auction provides higher revenue to mobile charging system when the number of drones increases.
In Fig.\ref{fig:fig4}, the horizontal axis of the experiment means the iteration of the proposed deep learning network training. The convergence of the deep learning networks during small number of iteration shows high adaptability to specific applications.
Fig.(\ref{fig4:payU5}-\ref{fig4:payU15}) show that the proposed deep learning-based auction can achieve stability in approximately 300 iterations. Fig.~\ref{fig4:payU51015} shows that the stability can be achieved much faster when the pre-trained network is used. This results mean that the proposed deep learning based auction has high adaptability; and thus it can be applied to the various environment with partial knowledge valuation distribution as presented in Sec.~\ref{sec:sec3}.
In Fig.\ref{fig:fig4}, the results show that the proposed auction guarantees the increased revenue of mobile charging station over SPA-0 and has a highly adaptive algorithm under partial knowledge distribution (a.k.a., not fully distributed knowledge).

\begin{table}[t]
        \caption{Revenue statistics (in Fig.~\ref{fig7:box})}
    \footnotesize
        \begin{center}
            \scalebox{1}{
            	\begin{tabular}{l|c|c|c}
                    \toprule
                      & SPA & $k = 1$ & $k = 3$\\
                    \midrule [1.0pt] 
                    Mean & 8.0536 & 8.0548 & 8.6032 \\
                    Top 25 percentile & 6.6003 &  6.6081 & 7.1733\\
                    Top 75 percentile & 9.0358 &  9.0372 & 9.3873 \\[0.2ex]
                    \bottomrule
            	\end{tabular}
        	}
        \label{tab:fig8pay}
        \end{center}
\end{table}

\begin{table}
        \caption{Revenue of mobile charging station (in Fig.~\ref{fig7:bar})}
    \footnotesize
        \begin{center}
            \scalebox{1}{
            	\begin{tabular}{l|c|c|c|c|c}
                    \toprule
                     Case & $(1)$ & $(2)$ & $(3)$ & $(4)$ & $(7)$\\
                    \midrule [1.0pt] 
                SPA & 7.5585 & 7.7175 & 6.1124 & 6.4550 & 5.7769\\
                $k=1$ & 7.7392 & 7.8891 & 6.2405 & 6.5808 & 5.9459\\
                $k=3$ & 7.9419 & 9.5227 & 8.8311 & 9.5005 & 9.9011\\[0.2ex]
                    \bottomrule
            	\end{tabular}
            	}
        \label{tab:fig8per}
        \end{center}
\end{table}
    
\begin{figure*}
  \centering
  \begin{subfigure}{.48\textwidth}
            \centering
            \includegraphics[width=1\textwidth]{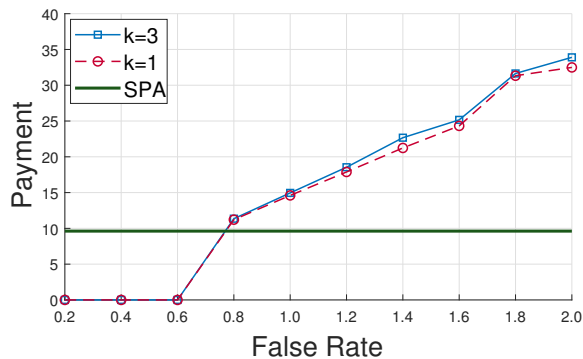}
            \caption{Payment changes due to false bidding.}
            \label{fig8:pay}
  \end{subfigure}
   \begin{subfigure}{.48\textwidth}
            \centering
            \includegraphics[width=1\textwidth]{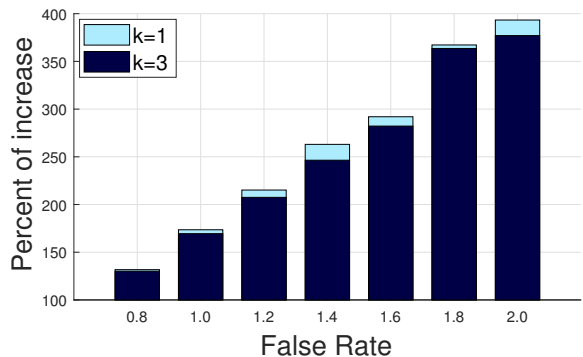}
            \caption{Increase of payment against SPA due to false bidding.}
            \label{fig8:percent}
  \end{subfigure}
  \caption{Payment comparison among drones.}
  \label{fig:fig8}
\end{figure*}

\BfPara{Statistical Analysis (Parameter $k$)}
In this section, we show the case where the penalty given to participant who bids a false bid. We confirm that the proposed method imposes penalty on the false bidder. 
In addition, the experimental results shows that how the penalty varies depending on $k$ values.
In Fig.~\ref{fig:fig4}, the effect of parameter $k$ can be observed while the number of drones varies. Fig.~\ref{fig:fig7} shows that the statistics analysis of revenue values for difference $k$ values in~\eqref{Eq5} and SPA-0 when the number of drones does not vary. The experiment results compare the average revenue, maximum revenue, minimum revenue, top 25 percentile, and top 75 percentile. The evaluation uses the deep learning networks when the $k$ values are $1$ and $3$. As $k$ increases, the gap between the average revenue of model and the average revenue of SPA-0 get larger as shown in Fig.~\ref{fig7:box} and Table.~\ref{tab:fig8pay}. When the proposed model is $k=1$, the revenue average is $8.0548$, similar to the revenue average of SPA-0. However, when the $k$ value of model is $3$, the revenue average is $8.6032$; and thus the model gets near $7$\% higher revenue average than SPA-0. When $k=1$, the gap between the proposed model and SPA-0 is near $0.008$ in terms of top $25$ percentile whereas the gap is about $0.57$, when $k=3$. In addition, in terms of top $75$ percentile, the revenue of $k=3$ model is about $0.3$ larger than $k=1$ model and SPA-0. This result shows that the proposed model with large $k$ takes higher revenue. Therefore, we can confirm that the revenue of mobile charging station declines in the order of $k=3$, $k=1$, and SPA-0. 
In Fig.~\ref{fig7:bar}, the graph shows the $10$ results of validation experiments, i.e., $10$ cases are considered in the validation experiments. The number on the $X$-axis in Fig.~\ref{fig7:bar} represents the indices of individual cases. The result stands for the revenue of mobile charging station via deep learning auction. We can confirm that the revenue with $k=1$ is always smaller than the one with $k=3$. The gap between the $k=1$ and $k=3$ models is about $0.2$ in Case $1$; and the Case $1$ is the minimum, whereas the maximum gap is about $5$ in Case $7$ as shown in Table~\ref{tab:fig8per}. The revenue with $k=1$ is larger than the SPA-0, but similar to SPA-0. The gap between the $k=1$ and SPA-0 is $0.13$ in Case $4$; and the Case is the minimum. The maximum gap is about $0.18$ in Case $1$. This experiments also show that the gap between the results by the two models with $k=1$/$k=3$ and the results of SPA-0 are not always equivalent. For example, the gap between $k=1$ and $k=3$ models is near $0.2$ in Case $1$m however near $1.7$ in Case $2$. This is due to the fact that the transformation depends on the weight of $\phi_i$; and thus the transformation via $\phi^{mononet}$ is not applied equally to the same bids. This means that if $b_{1}=2$ and also $b_{3}=2$, these two bids can be transformed differently. Therefore, the payment is not always equivalent. This means that the proposed deep learning auction adapts to the bid distribution at the time of the auction procedure, giving the mobile charging station high revenue. It can be seen that higher revenue is guaranteed by increasing the value of parameter $k$.

\begin{table}
        \caption{Payment of drone (in Fig.~\ref{fig8:pay})}
    \footnotesize
        \begin{center}
            \scalebox{1}{
            	\begin{tabular}{l|c|c|c|c}
                    \toprule
                        False Rate & $0.8$ & $1.2$ & $1.6$ & $2.0$\\
                    \midrule [1.0pt] 
                    SPA (6a) & 8.6177 & 8.6177 & 8.6177 & 8.6177 \\
                    $k=1$ (6a) & 11.2161 & 17.8814 & 24.3196 & 32.4915 \\
                    $k=3$ (6a) & 11.3513 & 18.5424 & 25.1623 & 33.8989 \\
                    $k=1$ (6b) & 130.15\% & 207.49\% & 282.20\% & 377.03\%\\
                    $k=3$ (6b) & 131.72\% & 215.16\% & 291.98\% & 393.36\%\\[0.2ex]
                    \bottomrule
            	\end{tabular}
            	}
        \label{tab:fig7bar}
        \end{center}
\end{table}

The proposed deep learning based auction algorithm has a strength in terms of giving penalty to false bidder.
In Fig.~\ref{fig:fig8}, experiment results present the payment of drone when the drone submits bid falsely (i.e., fake bid). This experiment conducts with the models of $k=1$ and $k=3$. The experiment assumes that the number of drones which participate in auction is $5$. The truth valuation of drone which submits the bid falsely is set to $15.9835$. The bid values of the other drones are generated by uniform distribution. This experiment uses the scenario where $5$ drones exist and one is with fake bid and the other four are with truthful bids. 
In Fig.~\ref{fig8:pay}, the second highest bid is set to $8.6177$ as shown in Table~\ref{tab:fig7bar}. This result shows that a drone cannot win the auction when it bids up to $0.2-0.8$ times larger than the true valuation $15.9835$. As a result, the drone is defeated in the auction due to false bid. On the other hand, when a drone submits bid as $1.2-2$ times larger than true valuation, the fake bid leads to win in auction. However, the fake bid increases the payment in the fake-bid drone. For example, if the drone submits near $40$ bid falsely in the SPA-based auction, the drone can only pay about $8.4$. However, the payment is $32.4915$ in the proposed auction with $k=1$. This means that the bidding of drone which falsely submit the bid for getting charging increases the payment. In addition, the payment increases when the $k$ value of models increases. Table~\ref{tab:fig7bar} also shows the payment increment while $k$ increases. The increased payment of the proposed model is at least $30$\% greater than that of the auction using SPA-0 as shown in Table~\ref{tab:fig7bar}. 
When the drone submits true valuation, the payment is just about $50$\% higher than the SPA-0 auction. However, if the bid is $1.2$ times larger than the true valuation, the payment is about $200$\% larger than the SPA-0 auction. This experiment shows that if drone submits false bid for winning the auction, the drone gets a loss in terms of the payment; and thus the loss let the drone avoid fake bidding. 

\begin{figure*}[t!]
   \centering
   \begin{subfigure}{.48\textwidth}
            \centering
            \includegraphics[width=1\textwidth]{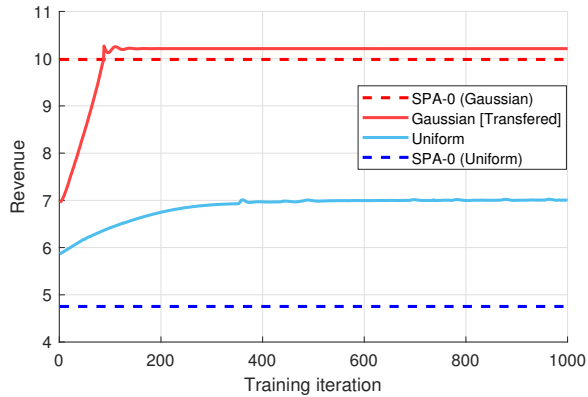}
            \caption{Transfer learning by changing bid distribution, $U=5$}
            \label{fig:transfer}
  \end{subfigure}
   \begin{subfigure}{.48\textwidth}
            \centering
            \includegraphics[width=1\textwidth]{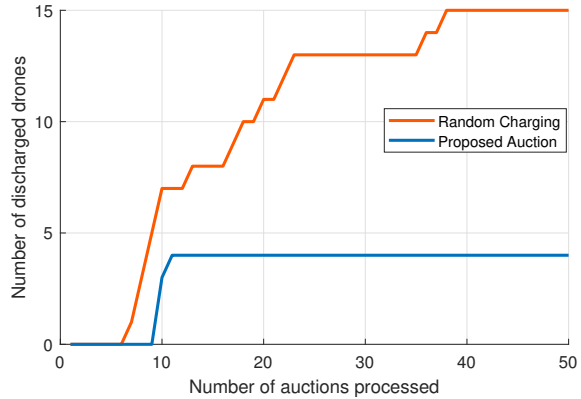}
            \caption{Number of discharging drones during the auction, $U=15$}
            \label{fig:battery}
  \end{subfigure}
  \caption{Performance improvements under various bid distributions.}
  \label{fig:revision}
\end{figure*}

Fig. (\ref{fig:transfer}) shows the proposed deep learning based auction can be trained through transfer learning when the distribution of bid values varies. The dotted lines are revenue when the SPA-0 is executed. The red and blue lines stand for the revenue when the proposed algorithm is used. The revenue is higher than SPA-0 as shown in previous experiments. The proposed model is stabilized with 400 training iterations if training starts from the initialized model. If the training starts from the trained model (i.e., transfer learning), the proposed model is stabilized with approximately 100 training iterations when the bid distribution varies. This experiment shows that the proposed model adapts to the change of the bid distribution and can provide reasonable results at various distribution. 
In Fig.~(\ref{fig:battery}), we consider the flight energy consumption of drones in this experiment as follows~\cite{pugliese2016modelling, zorbas2016optimal}. 
\begin{equation}
    \label{energy_consumption}
    E(t) = (\beta + \alpha z)\cdot t + P_{\max}\left(\frac{z}{s}\right)
\end{equation}
where $\alpha$ is a motor speed multiplier, $\beta$ is the minimum power needed to hover just over the ground (when altitude is almost zero), $z$ means the height at time step $t$, and $s$ is the speed of drone and $P_{\max}$ is the maximum power of motor to flight. Therefore, the term $P_{\max}(\frac{z}{s})$ refers to the power consumption needed to lift to height $h$ with speed $v$~\cite{pugliese2016modelling, zorbas2016optimal}. In this experiment, we set the values of $\alpha$ to $5.5$, $\beta$ to $15$, $z$ to $1m$, and $P_{\max}$ to $45$. Fig.~(\ref{fig:battery}) shows that the number of drones discharged from battery. In this experiment, we assumed that the charging service fully charges the battery of the drones which wins in the auction. The charging service is only for $1$ time slot. Therefore, the drones consume $65.5mAh$ per $1$ time step (1 hour) and recharge $1000mAh$ when recharged. 
In this experiment, we assume that $15$ drones exist and they want to constantly join the charging service scheduling. In Fig.~(\ref{fig:battery}), when the proposed deep learning auction is used to schedule charging services, $11$ of the $15$ drones can be charged without being discharged. This experiment shows that the proposed method can increase the drone flight time in multi-drone networks.

\begin{figure*}
  \centering
  \begin{subfigure}{.32\textwidth}
            \centering
            \includegraphics[width=1\textwidth]{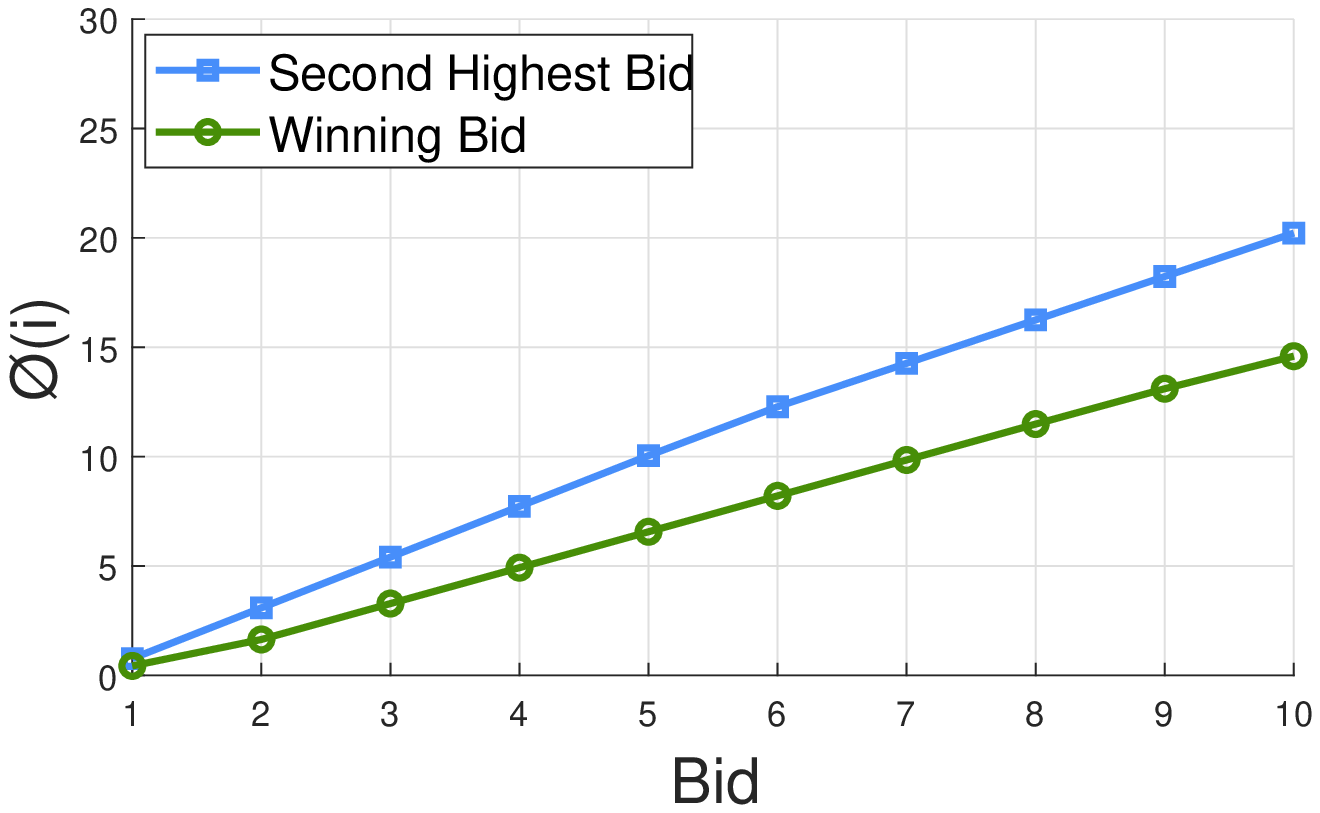}
            \caption{$b_i^\prime$ bid change, $U=5$}
            \label{fig5:linear5}
  \end{subfigure}
  \begin{subfigure}{.32\textwidth}
            \centering
            \includegraphics[width=1\textwidth]{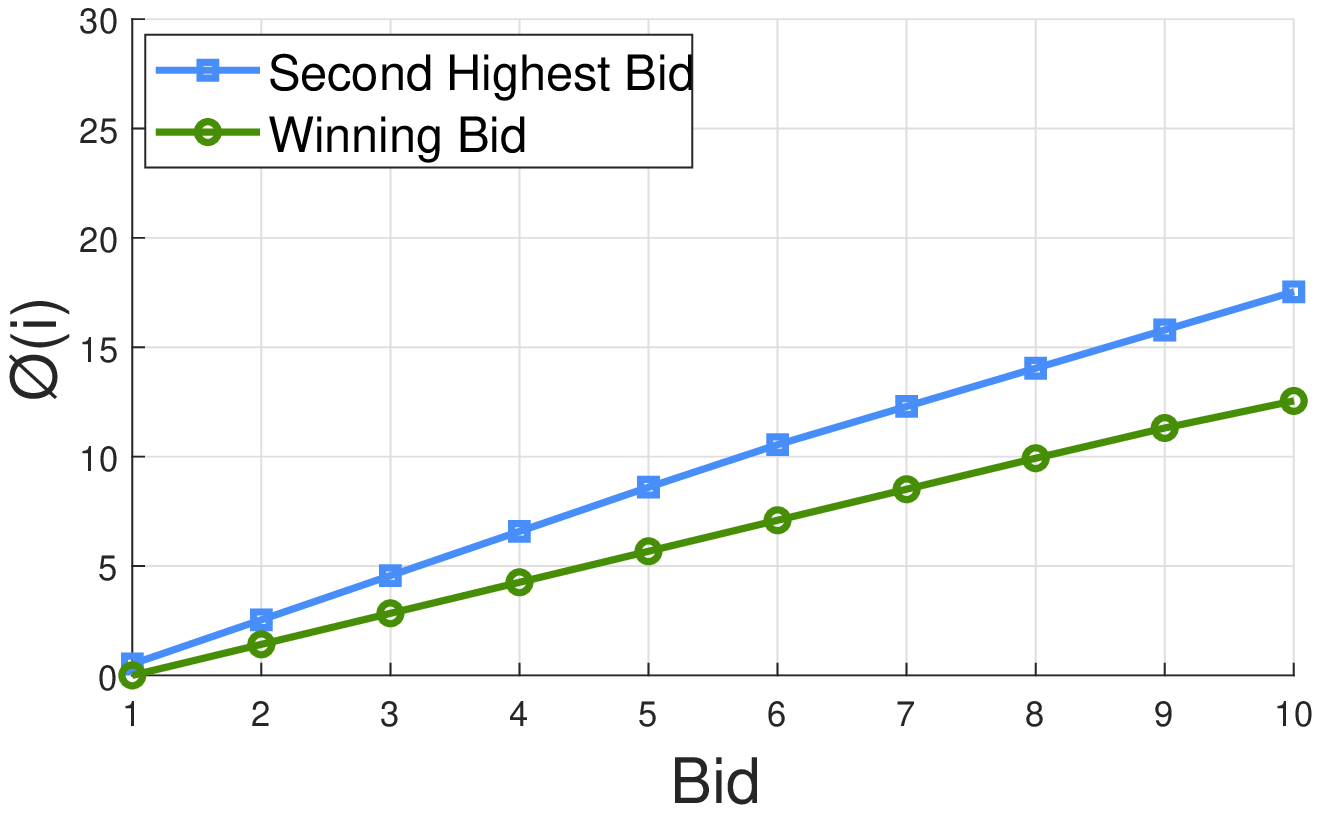}
            \caption{$b_i^\prime$ bid change, $U=10$}
            \label{fig5:linear10}
  \end{subfigure}
  \begin{subfigure}{.32\textwidth}
            \centering
            \includegraphics[width=1\textwidth]{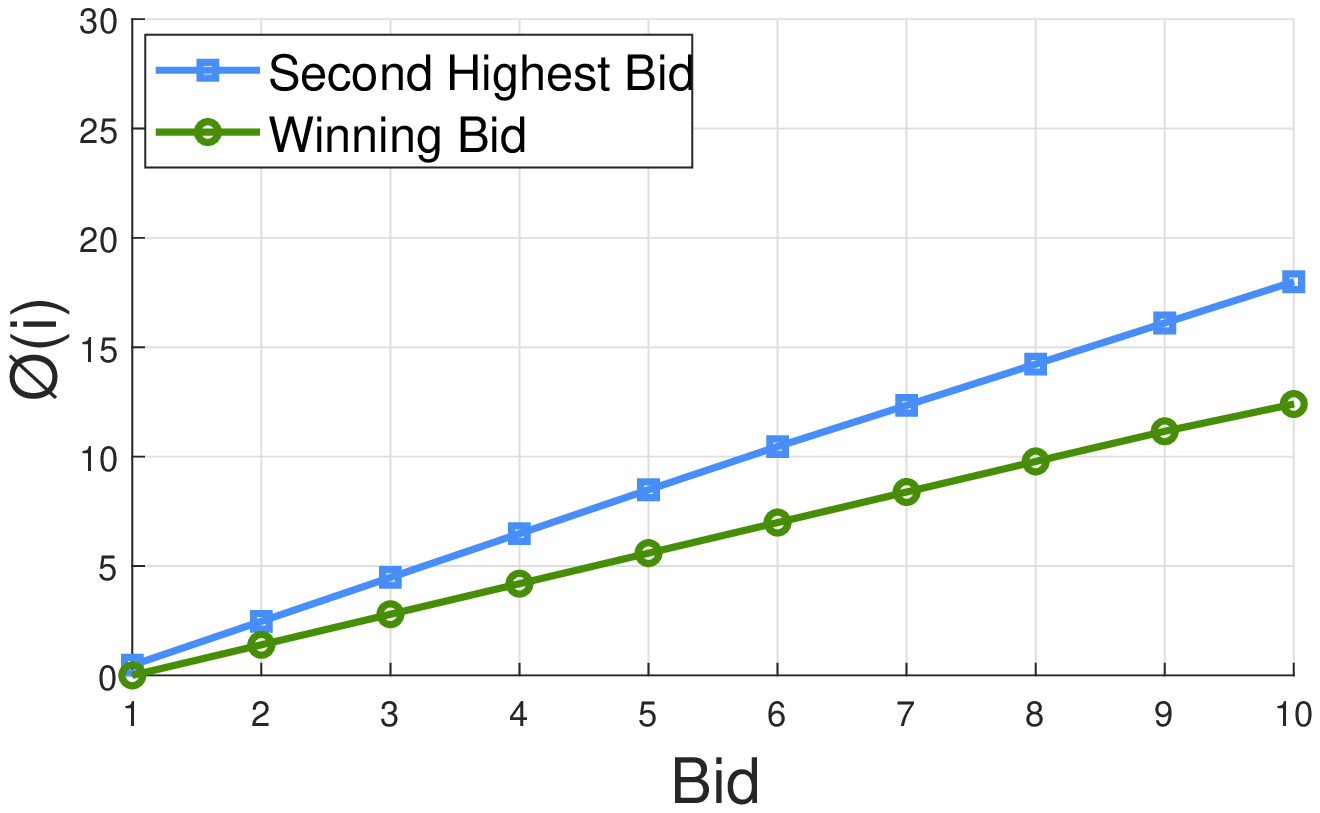}
            \caption{$b_i^\prime$ bid change, $U=15$}
            \label{fig5:linear15}
  \end{subfigure}
  \caption{$\phi_i(b_i)$ changes according to drone's valuation.}
  \label{fig:fig5}
\end{figure*}

\begin{figure*}
   \centering
   \begin{subfigure}{.32\textwidth}
            \centering
            \includegraphics[width=1\textwidth]{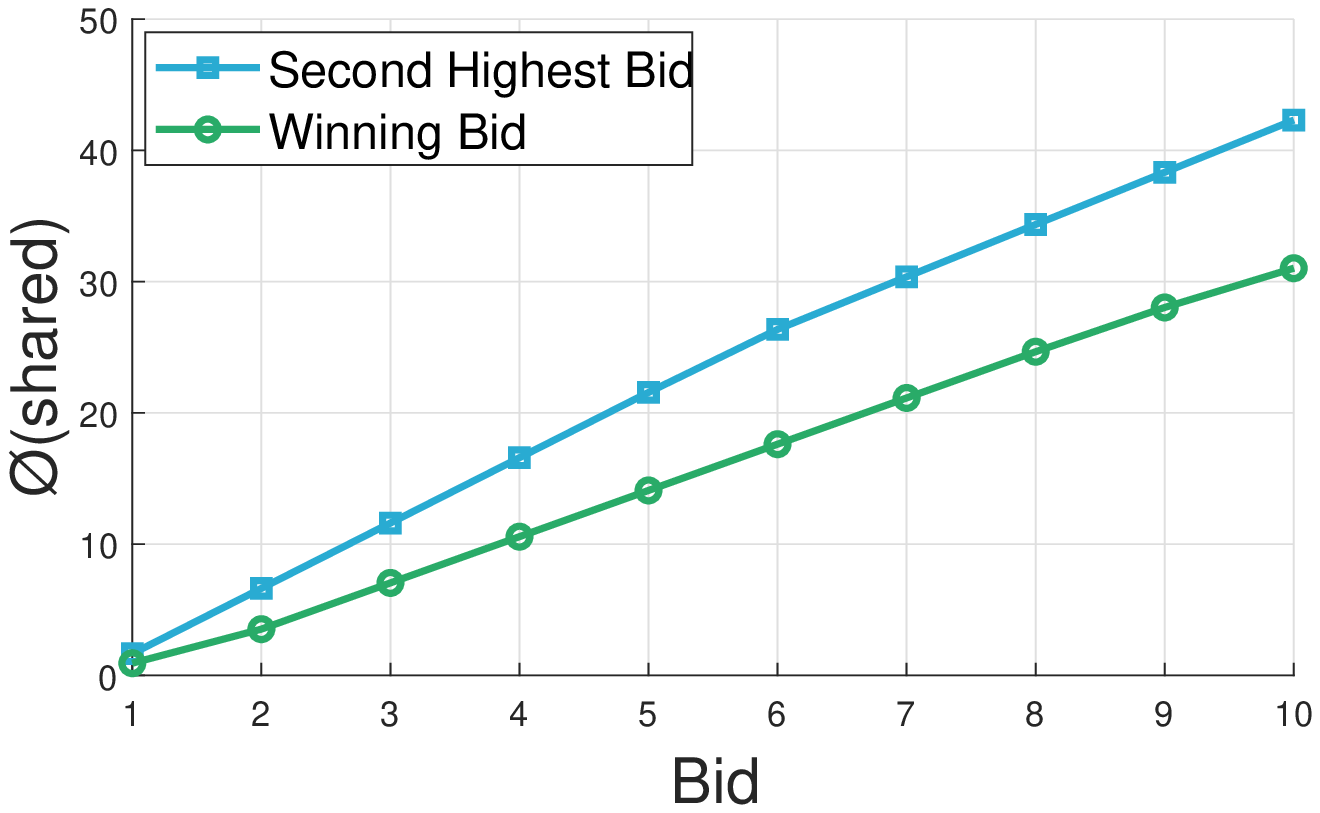}
            \caption{$\overline{b_i}$ bid change, $U=5$}
            \label{fig6:sharedU5}
  \end{subfigure}
   \begin{subfigure}{.32\textwidth}
            \centering
            \includegraphics[width=1\textwidth]{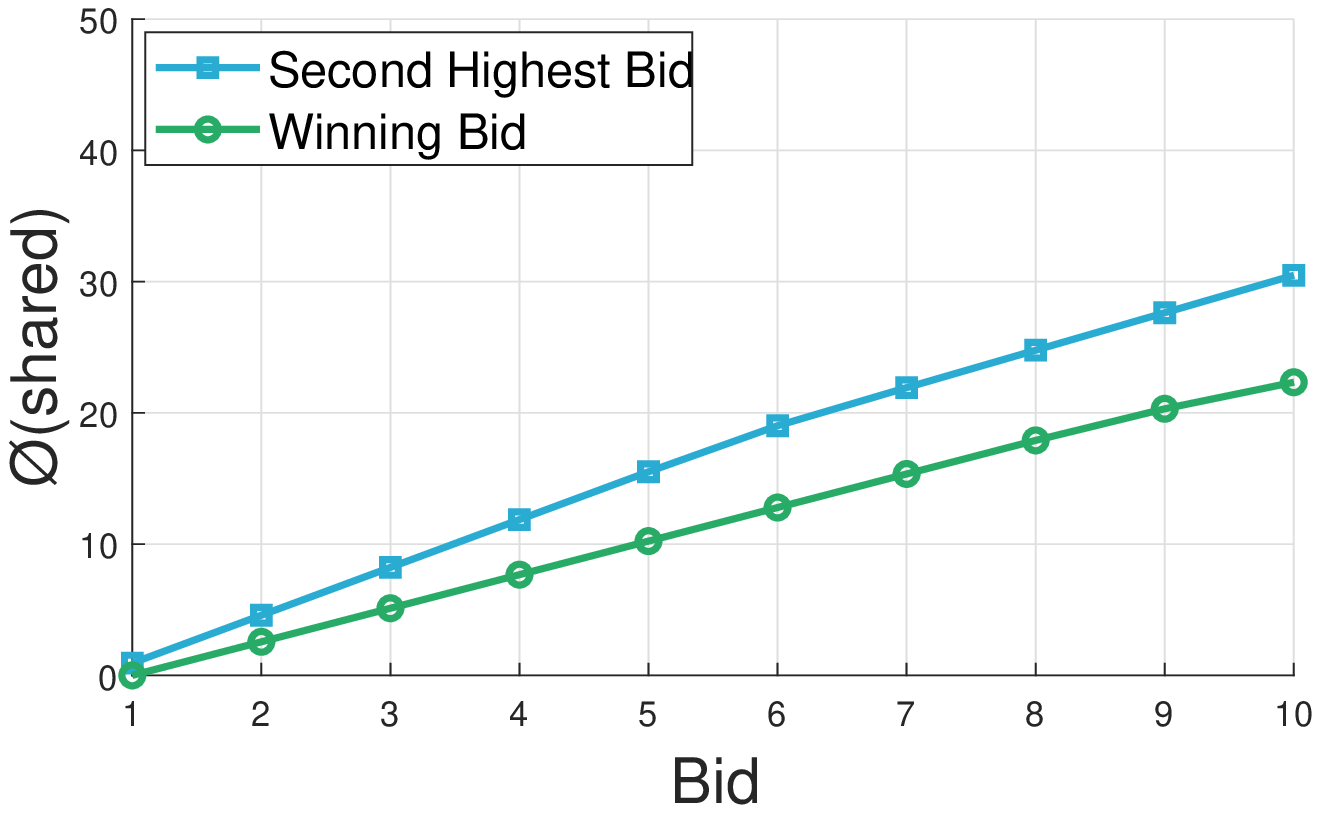}
            \caption{$\overline{b_i}$ bid change, $U=10$}
            \label{fig6:sharedU10}
  \end{subfigure}
   \begin{subfigure}{.32\textwidth}
            \centering
            \includegraphics[width=1\textwidth]{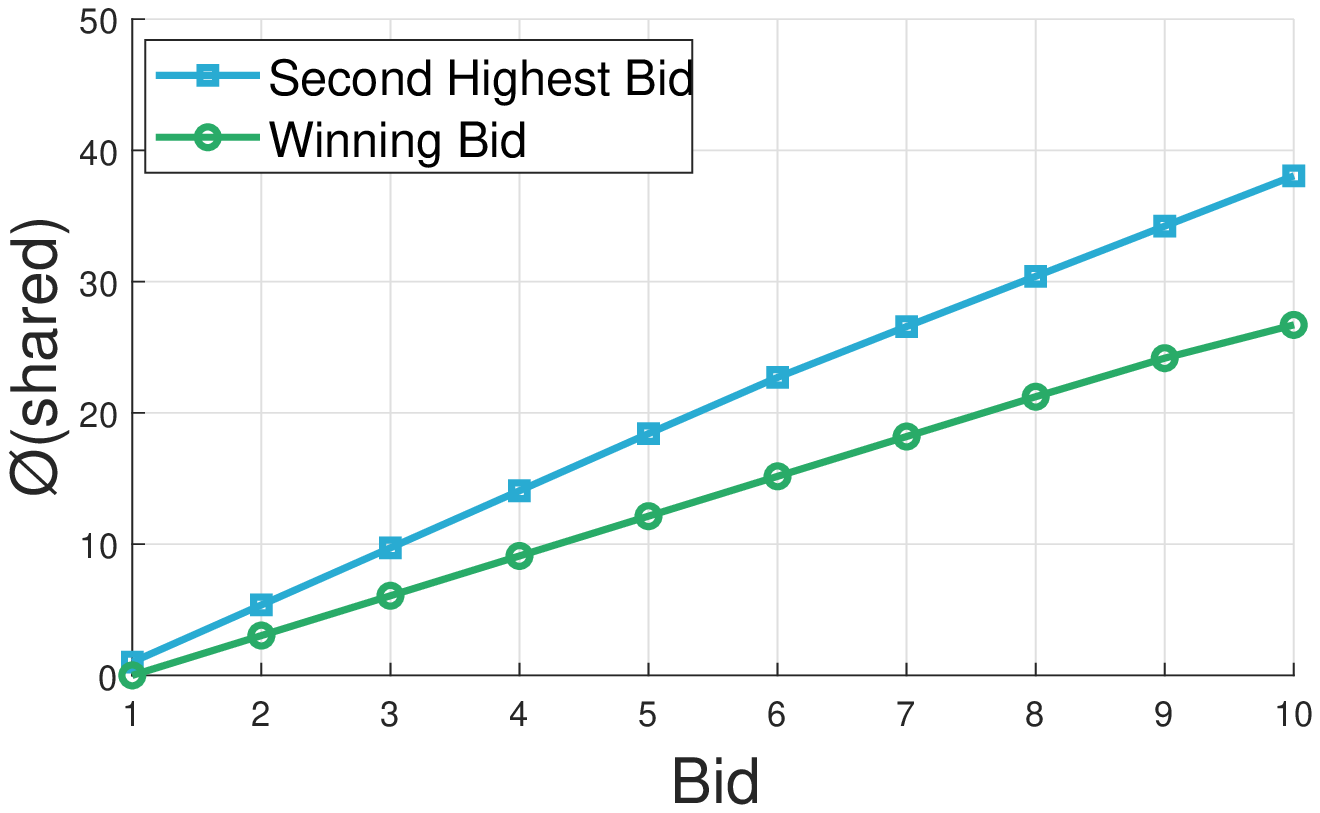}
            \caption{$\overline{b_i}$ bid change, $U=15$}
            \label{fig6:sharedU15}
  \end{subfigure}
  \caption{$\phi^{shared}(b_i^\prime)$ changes according to drone's valuation.}
  \label{fig:fig6}
\end{figure*}

\BfPara{Deep Learning Model Characteristics}
These experimental results explain why monotonic network is considered as the baseline structure of the proposed deep learning auction architecture. In the proposed auction, when bids are transformed to virtual values through the monotonic network, the order of bids must be maintained. For example, if the bid $b_1$ of user $u_1$ is larger than the bid $b_2$ of user $u_2$, converted $\overline{b_1}$ must be larger than $\overline{b_2}$. The corresponding experimental results show that the monotonic network performs the transformation that maintains the order.
Fig.~\ref{fig:fig5} shows the transformed value which is the result of virtual valuation function $\phi_i$ when the bid of the winner drone and the second highest bid increase from $1$ to $10$. The fixed weights of the $\phi_i$ were used when the winning bid and the second highest bid were transformed because the weights are continuously updated during deep learning training process. For this experiment, the networks are trained when the winning bid is $7.48292044$, the second highest bid is $4.75319804$, and other bids are generated along the uniform distribution. In Fig.~\ref{fig5:linear5}, the transformed value $p_i^\prime$ of the second highest bid is higher than the one of winning bid. This tendency can be shown in Fig.~\ref{fig5:linear10} and Fig.~\ref{fig5:linear15}. By comparing the Fig.~\ref{fig5:linear5} and Fig.~\ref{fig5:linear10}, it can be observed that the transformed bid decreases as the number of drones increases. However, through Fig.~\ref{fig5:linear10} and Fig.~\ref{fig5:linear15}, we can confirm that the transformed value $b_i^\prime$ is independent to the number of drones. Instead of the number of drones, the weights are affected by the allocation probability and payment of other bids because the loss function is configured based on the allocation rule and payment rule. 
Fig.~\ref{fig:fig5} also shows the $\phi_i$ network conducts non-decreasing monotonic transformation. Therefore, the $\phi_i$ network is able to replace $\phi$ function in Myerson auction because the $\phi_i$ network performs a monotonic transformation.

In Fig.~\ref{fig:fig6}, the changes of $\phi^{shared}(\phi_i(b_i))$ is presented when the bid of the winner drone and the second highest bid increase from $1$ to $10$. The fixed weights of the $\phi^{shared}$ were also used. The deep learning network is trained when the winning bid is $7.48292044$, the second highest bid is $4.75319804$, and other bids are generated along the uniform distribution similar to the evaluation for Fig.~\ref{fig:fig5}. It can be seen that the result of $\phi^{shared}$ is larger than the one of the $\phi_i$ shown in Fig.~\ref{fig:fig5} because the computation of $\phi^{shared}$ is conducted based on the result of the $\phi_i$ as well as the $\phi^{shared}$ does non-decreasing monotonic transformation. Through comparison of Fig.~\ref{fig6:sharedU5} and Fig.~\ref{fig6:sharedU10}, it can be seen that the result of $\phi^{shared}$ decreases when the number of drones increases, as shown in Fig.~\ref{fig:fig5}. However, in Fig.~\ref{fig6:sharedU15}, it can be observed that the result of $\phi^{shared}$ with $15$ drones becomes larger than the one of $\phi^{shared}$ with $10$ drones. Therefore, this experiment shows that the weights are affected by the allocation probability and payment and they are independent to the number of drones. The transformed bid $\overline{p_i}$ of the second highest bid is larger than the one of the winning bid. The tendency which is shown in Fig.~\ref{fig:fig5} is maintained.
The $\phi_i$ network has independent weights per input data. That is, if there are $5$ inputs, there are $5$ $\phi_i$ networks, and all learning networks have different weights. However, $\phi^{shared}$ network is just one network regardless of the number of inputs. If there are $5$ inputs, only one $\phi^{shared}$ network exists. Through Fig.~\ref{fig:fig5} and Fig.~\ref{fig:fig6}, it can be observed that the non-decreasing monotonic feature of $\phi_i$ networks and $\phi^{shared}$ are trained by the limited network structure and the loss function regardless of whether the weights are shared or not.

\section{Concluding Remarks and Future Work}\label{sec:sec5}
The proposed deep learning based auction is revenue optimal for mobile charging scheduling in distributed multi-drone networks. In this paper, the mobile charging scheduling problem is interpreted as auction problem where each drone bids its own valuation and then the charging station schedules drones based on it in terms of revenue-optimality. 
Through the proposed deep-learning based solution approach, the charging auction enables efficient scheduling by  automatically learning the required knowledge (i.e., bids distribution), which is required in conventional auction mechanisms. Therefore, environmental information is not required anymore in auction computation. This makes effective troubleshooting possible in distributed multi-drone networks. The proposed algorithm only requires payment and allocation probabilities by the multi-drones. The loss function in deep learning computation is an important factor that allows the proposed auction to be constructed based on environment independent information. 
As verified via software prototype based performance evaluation, following facts are observed: 
    (i) guaranteeing optimal revenue in terms of individual rationality and dominant strategy incentive compatibility,
    (ii) limiting the false bids of drones by increasing the payment to the false-bid drones,
    and
    (iii) enabling a revenue optimal auction to be constructed without complex prior knowledge, i.e., bids distribution.
    
As future research directions, advanced auction mechanism designs with multiple mobile charging stations are worthy to consider. In this case, the problem can be formulated with multi-item auction and then the corresponding mathematical formulation, verification, and analysis are desired.
Furthermore, the proposed deep learning-based auction mechanisms can be advantageous in various applications. For example, visual attention is considerable because it can be reformulated as resource allocation~\cite{zhang2016detection,zhang2017co, han2006unsupervised, han2015background}.

\bibliographystyle{IEEEtran}  
\bibliography{ref_tvt}


\newpage

\begin{IEEEbiography}[{\includegraphics[width=1in,height=1.25in,clip]{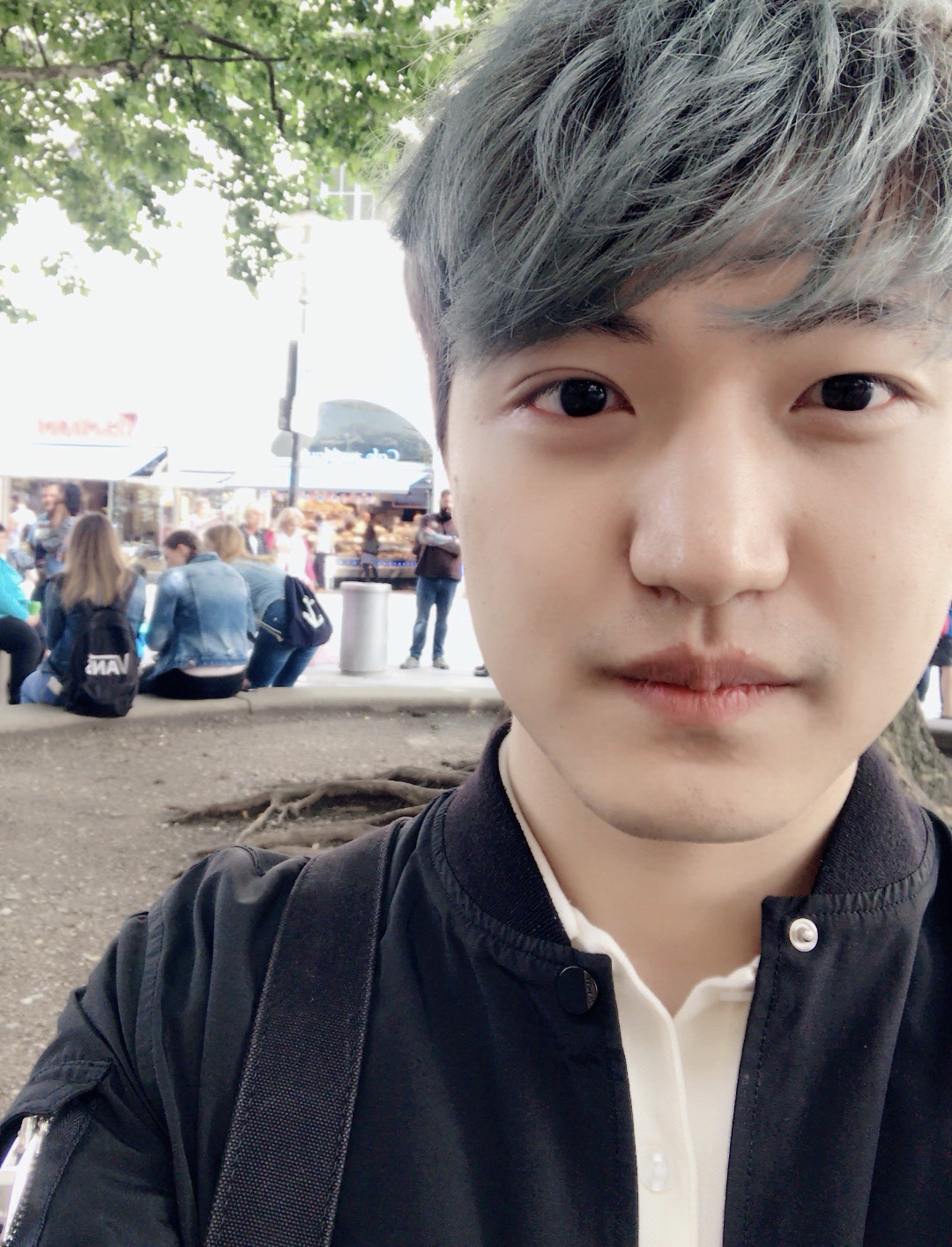}}]{MyungJae Shin}
is currently an M.S. Student in Computer Sciene and Engineering, Chung-Ang University (CAU), Seoul, Korea. He received his B.S. in computer science and engineering from CAU, Seoul, Korea, with the second highest honor from the CAU College of Engineering. His research interests are in various econometric theories and their deep-learning based computational solutions. He was a recipient of the National Science \& Technology Scholarship (2016--2017).
\end{IEEEbiography}

\begin{IEEEbiography}[{\includegraphics[width=1in,height=1.25in,clip]{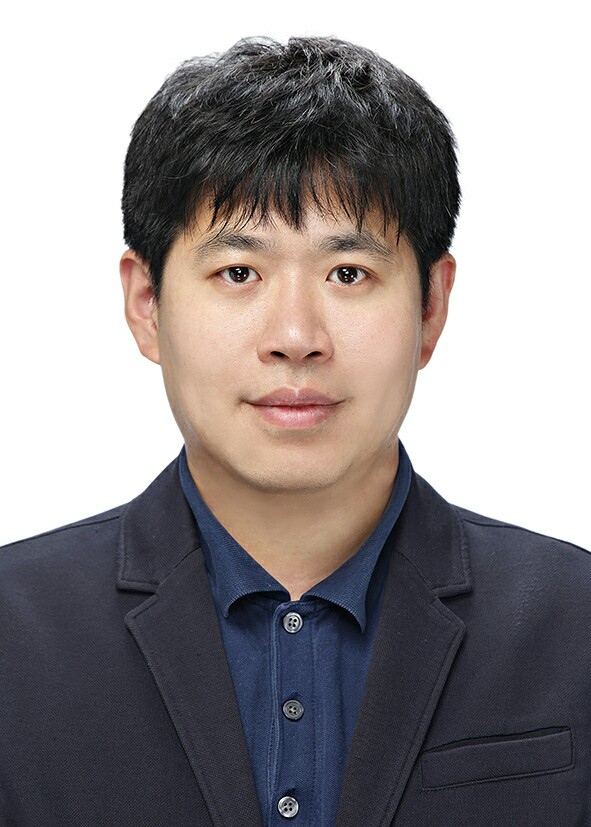}}]{Joongheon Kim}
(M'06--SM'18) is currently an assistant professor with Chung-Ang University School of Computer Science and Engineering, Seoul, Korea, since 2016. He received his B.S. (2004) and M.S. (2006) in computer science and engineering from Korea University, Seoul, Korea; and his Ph.D. (2014) in computer science from the University of Southern California (USC), Los Angeles, CA, USA. In industry, he worked for LG Electronics (Seoul, Korea, 2006--2009), InterDigital (San Diego, CA, USA, 2012), and Intel Corporation (Santa Clara, CA, USA, 2013--2016). 

He is a senior member of the IEEE. He was a recipient of the Annenberg Graduate Fellowship with his Ph.D. admission from USC (2009) and the Haedong Young Scholar Award (2018) which is for recognizing a young Korean researcher under the age of 40 who has made outstanding scholarly contributions to communications and information sciences research.
\end{IEEEbiography}

\begin{IEEEbiography}[{\includegraphics[width=1in,height=1.25in,clip]{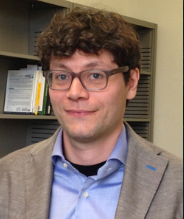}}]{Marco Levorato} joined the Computer Science department at University of California, Irvine in August 2013. Between 2010 and 2012, He was a post-doctoral researcher with a joint affiliation at Stanford and the University of Southern California working with prof. Andrea Goldsmith and prof. Urbashi Mitra. From January to August 2013, he was an Access post-doctoral affiliate at the Access center, Royal Institute of Technology, Stockholm. He is a member of the ACM, IEEE and IEEE Comsoc society. His research interests are focused on next-generation wireless networks, autonomous systems, Internet of Things, e-health and stochastic control. He has co-authored over 100 technical articles on these topics, including the paper that has received the best paper award at IEEE GLOBECOM (2012). He completed the PhD in Electrical Engineering at the University of Padova, Italy, in 2009. He obtained the B.S. and M.S. in Electrical Engineering summa cum laude at the University of Ferrara, Italy in 2005 and 2003, respectively. In 2016, he received the UC Hellman Foundation Award for his research on Smart City IoT infrastructures.

\end{IEEEbiography}

\end{document}